\documentclass{svjour2}                    % onecolumn
\smartqed  % flush right qed marks, e.g. at end of proof
\usepackage{graphicx}
\begin{document}

%%% Begun 3/9; Last drafts before trip, 5/21 and 6/10; to AIP format 7/19/08
\def\HIDE#1 {{}}
\def\OMIT#1 {{}}
\def\rem#1 {{}}
\def\MEMO#1 {}
\def\LATER#1 {}
\def\TEMP#1 {}
%%%%%%%%%%%%%%%%%%%%
\newcommand{\beq}{\begin{equation}}
\newcommand{\eeq}{\end{equation}}
\newcommand{\eqr}[1]{(\ref{#1})}
\newcommand{\half}{\frac{1}{2}} 
\newcommand{\la}{{\langle}}
\newcommand{\ra}{{\rangle}}
\newcommand {\HH} {\mathcal{H}}  %% same as \Ham
\newcommand {\ep}{\epsilon}
\newcommand {\epdot} {\dot{\ep}}

\newcommand{\xaxis}{{\bf \hat{x}}}
\newcommand{\yaxis}{{\bf \hat{y}}}
\newcommand{\zaxis}{{\bf \hat{z}}}
\newcommand{\ssigma}{\underline{\sigma}}
\newcommand{\rr}{{\bf r}}
\newcommand{\jj}{{\bf j}}
\newcommand{\vv}{{\bf v}}
\newcommand{\TT}{{\bf T}}
\newcommand{\FF}{{\bf F}}
\newcommand{\zz}{{\bf z}}
\newcommand{\EE}{{\bf E}}
\newcommand{\rcargo}{{r_{\rm cargo}}}
\newcommand{\Ftip}{{F_{\rm tip}}}
\newcommand{\Fdrag}{{F_{\rm drag}}}

\title{Possible origins of macroscopic left-right asymmetry
in organisms
}
\subtitle{subtitle here}

\titlerunning{Running title}        

\author{Christopher L. Henley}

%%% *** Motor proteins (myosin, kinesin dynein) 87.16.Nn
%%% *** Filaments, microtubules, their networks, and 
%%%    supramolecular assemblies 87.16.Ka 	
%%% *** Development and growth 87.19.lx 	
%%% *** (Cell) Growth and division 87.17.Ee 	
%%% * subcellular processes: Transport, incl. channels, pores, 
%%%       lateral diffusion 87.16.dp 	
%%% Pseudopods, lamellipods, cilia, and flagella 87.16.Qp 	
%%% subcellular processes: analytical theories 87.16.ad 	
%%% structure of biomolecules (incl. their chirality) 87.15.B-
%%% Cytoskeleton, 87.16.Ln
%%% Biol physics -- Analytical theories 87.15.ad   (gen. theory and sim 87.15A-)
%%% Cell locomotion, chemotaxis 87.17.Jj 	
%%% (Higher organisms) Pattern formation: activity and anatomic 87.19.lp 	

\institute{Christopher L. Henley \at
{Department of Physics, 
Cornell University, Ithaca, New York 14853-2501, USA} \\
              \\
              Tel.: +1-607-255-5056\\
              Fax: +1-607-255-6428 \\
              \email{clh@ccmr.cornell.edu}        
}

\date{Received: date / Accepted: date}

\institute{F. Author \at
              first address \\
              Tel.: +123-45-678910\\
              Fax: +123-45-678910\\
              \email{fauthor@example.com}           %  \\
           \and
           S. Author \at
              second address
}

\date{Received: date / Accepted: date}

\begin{abstract}
I consider the microscopic mechanisms by which
a particular left-right (L/R) asymmetry 
is generated at the organism level from the microscopic handedness of
cytoskeletal molecules.
In light of a fundamental symmetry principle,
the typical pattern-formation mechanisms 
of diffusion plus regulation 
cannot implement the ``right-hand rule''; at the
microscopic level, the cell's cytoskeleton of chiral filaments
seems always to be involved, usually in collective states 
dirven by polymerization forces or molecular motors.
It seems particularly easy for handedness to emerge in
a shear or rotation in the background of an effectively  two-dimensional
system, such as the cell membrane or a layer of cells, 
as this requires no pre-existing axis apart from the layer normal.
I detail a scenario involving actin/myosin layers in snails 
and in {\it C. elegans}, and also one about the microtubule
layer in plant cells.  I also survey the other examples 
that I am aware of, such as the emergence of handednes
such as the emergence of handednes in neurons, in eukaryote 
cell motility, and in non-flagellated bacteria.
\end{abstract}

\maketitle

%%%%%%%%%%%%%%%%%%%%%%%%%%%%%%%%%%%%%%%%%%%%
%% MAINMATTER
%%%%%%%%%%%%%%%%%%%%%%%%%%%%%%%%%%%%%%%%%%%%

\section{Introduction}
\label{sec:intro}

Most animals -- and many plants --  
have systematic chiral asymmetries,  uniform
throughout their species.
Yet pattern-formation in development is based on
reaction-diffusion equations -- where ``reaction''
includes the rate equations of gene regulation
and expression -- and perhaps on elasticity.
But diffusion and elasticity, being described physically
by symmetric tensors, are not chiral; the chiral
information must somehow be brought from the molecular
scale to the macroscopic scale.  It is reminiscent
of Feynman's fantasy of how one could communicate the 
absolute definitions of left and right to extraterrestrials
purely by radio: they can perform particle physics experiments
that access the inherent left-right asymmetry of physics
on the subatomic scale.  To achieve a systematic handedness,
every embryo must analogously perform an experiment that
accesses the left-right asymmetry of biochemistry at the
molecular scale.
Embryologists have recognized this as a fundamental question
over the past 
25 years~\cite{brown-wolpert,wood-review,levin-LR-review,wood-review-2005}.

Why is handedness an attractive topic for theoretical modeling,
and why should it be claimed by physics?
In the first place, unlike most pattern-forming
instabilities in development,
handedness is a choice between {\it equivalent} outcomes 
and is hence more clearcut to study~\cite{palmer}.
In other words it is a {\it symmetry breaking}, akin to that in the
Ising model of statistical physics.
Being binary also makes it amenable to typical biological
experiments, where the outcomes are classified qualitatively.
And {\it symmetry} puts constraints on
the conceivable mechanisms, giving the theorist a rare 
opportunity to make {\it a priori} statements about biology.

Secondly, it appears that the mechanisms involve {\it physics}
in a central fashion: forces and motions in the cytoskeleton.
Now, most stories in biology are about {\it signaling} and it is
natural that biologists look for signal-based mechanisms.  
Signal-based pattern formation requires two ingredients:
(1) gene regulation or other reactions, and
(2) transport of signaling molecules by diffusion, or by
active transport that similarly just depends on proximity.
The diffusion ingredient is responsible for the spatial patterning, 
but {\it it cannot  produce an L/R spatial patterning}, since
diffusion does not distinguish handedness 
(see Sec.~\ref{sec:diffusion-LR-no}).
The actual mechanisms, insofar as we know them,
involve literal {\it forces} or {\it torques} 
exerted by molecular motors, acting along the 
{\it cytoskeleton} of stiff, semi-macroscopic fibers,
and in this sense the topic belongs to mechanics.

This paper (superseding Ref.~\cite{LR-Landau}) is 
a review of sorts, but not mainly of theoretical results 
(which are few).  Instead, it aims to survey diverse systems
manifesting left/right asymmetries, with the aim of finding
commonalities and classifying the differences usefully.
The hope is that -- since the mechanism is strongly
constrained by symmetry -- one can exhaustively categorize
the models and that this catalog will be helpful in finding the
right model for each system.  What is actually done here is
to lay out the facts for the most promising systems and attempt
to imagine all the plausible models for each, as groundwork
for future modeling work (or as motivation for new
experiments needed before we can estimate basic parameters 
in the models).  The specific hypothetical proposals sketched
are my own, except where previous work is mentioned and cited.
What is probably most novel here is to consider how the handedness 
may realistically be implemented at the molecular level in
specific cases; the more physical ideas in past discussions
have mostly been at the more macroscopic level of mechanics.
The most penetrating existing thoughts about mechanical (but not 
molecular) mechanisms are found in Wood's review~\cite{wood-review}.

Note that the question of this paper is quite separate 
from another one that has long attracted physicists,
the way organic molecules originally
broke left-right symmetry in prebiotic evolution~\cite{frank53,avertisov-molecules}
%%% garcia-bellido}.

\subsection{Outline of paper and list of examples}
\label{sec:outline-list}

The paper begins (the rest of Section~\ref{sec:intro})
by laying out the necessary properties of left/right mechanisms
(along with some assumptions I will adopt).  
Furthermore it lays out the important categories:
spontaneous instability (=symmetry breaking) versus 
proportionate response (Subsec.~\ref{sec:SSB}),
cell level versus multicellular level (Subsec.~\ref{sec:classify-levels}),
screw versus anchoring-molecule mechanisms (Subsec.~\ref{sec:classify-screw}),
and single-unit versus collective mechanisms.

After that, the sections run through the major experimental
examples of L/R asymmetry, which I will now preview in order.

\subsubsection{Example: vertebrates and cilia-driven flow}
\label{sec:cilia-flow}

The best-known left-right asymmetry is the internal organs 
(viscera) of vertebrates~\cite{cooke-review,hirokawa-review} --
among humans,~\cite{mcmanus-RHLH} just $\sim~10^{-4}$
of us have mirror-reversed insides (so that the heart
is on the right side).  Most vertebrate studies 
use one of four model systems:
mouse, chick, {\it Xenopus} frog, or zebrafish.

In 1998, it was observed through video imaging that
(at one stage) the surface of the mouse embryo has
cells with cilia whose tips move {\it circularly}
in a clockwise sense,~\cite{nonaka-cilia,hamada-review}
which breaks the symmetry 
sufficient to drive a leftwards flow 
in the surrounding fluid~\cite{nodecilia-fluid-dyn}.
If some (yet undiscovered) chemical signal is emitted,
it will be preferentially carried to the left and
distinguish that side, so the cilia-driven flow is
clearly {\it sufficient} to trigger handedess. 
Indeed, in one experiment that reversed the flow in a 
few embryos, they subsequently developed in a 
mirror reversed form~\cite{nonaka-hamada-reverse}.

This is the accepted picture, but subsequent studies
have complicated it: e.g., in the zebrafish, 
the cilia are on the internal
surface of an enclosed sphere, called Kupfer's vesicle,
and a different version of the fluid dynamics story will
be called for. 
More importantly, left/right asymmetries are reported in 
vertebrates long {\it before} the cilia start moving, 
in the form of a polarization of the gap junctions
connecting cells, leading to 
ion flow~\cite{levin-LR-review,levin-ionflow-review}.
Thus cilia do not {\it necessarily} determine handedness;
indeed a non-cilia mechanism is claimed to determine the 
body asymmetry in the case of {\it Xenopus} 
frogs~\cite{levin-xenopus,levin-aw}.

In this paper, I will skip the often-told story of vertebrates 
and focus on other examples for which there microscopic
physical mechanisms have not even been proposed, let alone accepted.

\subsubsection{Example: molluscs and {\it C. elegans}}

Another long-known example is the chirality of mollusc shells,
which has been studied in many species, most of which are
dominantly right-handed (reversal frequency $\sim~10^{-4}$).
And even the tiny {\it C. elegans} nematode ``worm'' has 
L/R asymmetry~\cite{wood-review,wood-review-2005,wood-orig,poole-hobert},
%% (they develop in an identical, and known, pattern),
%% and every embryo repeats exactly the same sequence of divisions.
most significantly in the left and right copies of a
certain chemosensing neuron.

In both molluscs and C. elegans, 
a twist in the positioning of cells during the
first three divisions determines the embryo's handedness.
This seems to be due to the chiral dynamics 
of a actin-myosin layer associated with the membrane.
In sec.~\ref{sec:actomyosin}, I will walk in some 
detail through three levels of description, 
proposing some physics on each level.
This is the most important section of this paper.
These stories are not quantitative, but they do lead
to clear predictions of the {\it sign} of handedness
in various phenomena.

\subsubsection{Example: cortical microtubules in plant cells}

There is a surprisingly similar story of handedness in plants
(Sec.~\ref{sec:mt-plants}), often manifested by
roots and shoots taking helical paths 
-- e.g. the twining of vines -- 
with a species-dependent handedness.
At the microscopic level, plant cells develop an array of 
parallel microtubules along the membrane,  
which has some twist around the cylindrical cell's long  axis. 
This is believed to determine the sense of
macroscopic helical behaviors, 
A couple of mechanisms.
will be sketched in Sec.~\ref{sec:mt-plants}
for the way that the array gets aligned helically.

\subsubsection{Example: nerve cells and brain handedness}

Section~\ref{sec:actin-motility} gathers several examples
related to motility and/or intercellular forces. The most
interesting of these is the human brain:
we show right hand dominance as a side effect 
of left brain dominance~\cite{mcmanus-LHRH,sun-brain-review}.
The brain handedness probably has a different mechanism from the viscera, 
since it is observed to be {\it independent} of the handedness
of internal organs~\cite{levin-LR-review,cooke-review,mcmanus-RHLH},
\OMIT{This independence is also confirmed in frogs~\cite{malashichev-review}.}
%%%%%%%%%%%%%%%%%%
and is less reliable by a factor of 1000
(1/10 of us are right-brain-dominant).
It is unclear whether the brain asymmetry other animals is
related to ours~\cite{malashichev-review,malashichev-book};
in the case of zebrafish, it seems to just follow the handedness
of internal organs~\cite{halpern-LR}.

To relate the macroscopic level to lower ones,
we note that nerve tissue forms by migration of 
nerve cells and by the growth
of very long axons (or dendrites). Growing axons indeed are observed 
to turn chirally, reminiscent of plant roots~\cite{sec:neurites}.

Both migration and growth dynamics depend on the cell extending
pseudopods, that are driven by actin polymerization forces in
the bulk.  Therefore, I outlined a (highly speculative) 
mechanism based on actin polymerization (Sec.~\ref{sec:actin-pitch}).
I have lumped in the recent discovery of chiral asymmetries in 
the motility of (non-nerve) mammalian cells (Sec.~\ref{sec:mammalian-channels})
and in the gut tissues of flies (Sec.~\ref{sec:drosophila}). 

\subsubsection{Bacteria}

For completeness and because their handedness is to some degree
understood, I include bacteria in the story (Sec.~\ref{sec:bacteria}).
The interesting case is when handedness is determined by 
the bacterial cytoskeleton (Sec.~\ref{sec:bacteria-MreB}).

\subsection{Cell level and multicellular level stories}
\label{sec:classify-levels}
%%% \label{sec:classify}

Any example of L/R specification requires explanations
on (at least) two levels. 
The {\it cell level} story takes us from molecules to
to handed behavior at the level of the entire cell.  
This typically emerges from some collective state of the 
cytoskeleton, which can be modeled in the framework of 
non-equilibrium statistical mechanics.
The {\it multicellular level} story starts from the cell behavior 
and explains how this specifies a macroscopic handedness in the 
whole embryo (or plant), typically in the framework of solid mechanics.
A sort of multiscale modeling is called for, meaning we 
not only have a theory for each level, but also can numerically
compute the parameters governing one level from the parameters
of the level below it.

In the case of the circularly moving cilia that affect handedness
in the mouse and other vertebrates (Sec.~\ref{sec:cilia-flow})
the cell level story
is merely the growth of one cilium containing dynein molecules
connecting microtubule pairs in a clockwise 
direction, as viewed from above.
The multicellular level  story involves not solid but
fluid mechanics in this case~\cite{nodecilia-fluid-dyn}
(the exterior fluid flow driven by cilia).

\subsection{Symmetry constraints and basic assumptions}
\label{sec:question-assumptions}

In this subsection, I want to identify necessary ingredients
in a left/right mechanism; I argue that basic symmetry
and  statistical mechanics put {\it a priori} constraints
on what  can work.

\subsubsection{Pre-existing axes and two-dimensionality}
\label{sec:x-z-cross}

Any organism can develop mutually orthogonal axes $\zaxis$, $\xaxis$, and
$\yaxis$, through a combination of reaction and diffusion by
chemical signals.  In an embryo, $\xaxis$ might represent
the {\it anterior/posterior} (A/P) axis
and $\zaxis$ might represent the {\it dorsal/ventral} (D/V)  axis;
within a single cell, $\xaxis$ might represent a direction of polarization
and $\zaxis$ might represent the normal to the membrane.
The problem is how to ensure the third axis satisfies a ``right-hand rule'' 
   \beq
     \yaxis = \zaxis \times \xaxis
   \label{eq:rhrule}
   \eeq
rather than its opposite.

A key observation~\cite{LR-Landau} is that symmetry requires that 
{\it each of the three symbols} on  the right-hand side of 
\eqr{eq:rhrule} has {\it a specific physical (biological) correlate} 
which is part of the mechanism: i.e., two kinds of preexisting polarization
representing $\xaxis$ and $\zaxis$, and something chiral 
representing the ``$\times$'' in \eqr{eq:rhrule}.
The above symmetry principle applies at {\it each} of the levels
distinguished in Sec.~\ref{sec:classify-levels}.
At the bottom level,
the chiral element is obviously the microscopic chirality 
of molecules, under genetic control, which can enter in a couple
of ways (see Sec.~\ref{sec:classify-screw}).

The above observation may appear trivial, but it has a
less trivial corollary.  
The commonest situation in which  {\it two} kinds of pre-existing 
polarization are present 
\OMIT{representing $\xaxis$ and $\zaxis$ in \eqr{eq:rhrule},}
is a {\it planar} geometry such that ``up'' and ``down'' directions 
are  inequivalent, so the $\zaxis$ is the unit normal to that plane.
On the cell level such a plane is the {\it membrane}; 
on the organism level, it means the mechanism should
operate at a stage during development when the embryo 
is still roughly planar in shape.
It seems harder to implement handedness inside a three-dimensional bulk,
and I do not know of a plausible such mechanism in any system.

\subsubsection{``Hall effect''in diffusion?}
\label{sec:diffusion-LR-no}

Development is governed normally by {\it reactions} and {\it diffusion}.
Here ``reactions'' includes regulation of DNA transcription and translation
to proteins.
Diffusion usually means a chemical species is produced by some cells 
and diffuses to others (either in the intercellular medium,
or via a network of gap junctions linking neighboring cells).
A nonzero net transport emerges statistically 
from local steps that appear rather random.  
%%%%%%%%%%%%%%%%%%%%%%

It would be amusing (but wrong) if the chiral bias 
could emerge via the
diffusion analog of the Hall effect (of electrical conduction
in magnetic fields). The usual Fick's law for the diffusion 
current reads
   \beq
         \jj = - D \nabla \rho
   \eeq
where $\rho$ is the number density.
In the presence of chiral scatterers, do we get another term 
proportional to $\hat z \times \nabla \rho$,
for some axis $\hat z$, so that the current deviates
by some regular angle from the concentration gradient?
The answer is no: diffusion in 
equilibrium satisfies time reversal symmetry, whereas
such a ``Hall effect'' violates time reversal symmetry.
I suspect it can be proven that the ``Hall'' bias is
negligible so long as the work done during a velocity 
correlation time is small compared to a thermal energy
$k_B T$; it would be an interesting challenge to turn 
that into a theorem.  

Such a theorem need not
apply to cytoskeletal dynamics,
since the motions of motor molecules and the 
polymerization of fibers are far from equilibrium
processes. Furthermore, the motion of a motor or the
growth state of a fiber tends to persist for many
steps, ensuring a long velocity correlation time.
For these reasons, I will consider only cytoskeletal-based
mechanisms. That seems to be the expectation of biologists
also, presumably because cytoskeletal dynamics is the 
normal way to generate asymmetries such a polarization in
a cell; also, in all particular examples for which we have partial
experimental information, the mechanism does seem to be
cytoskeletal.

\subsection{Filament screw or membrane-anchoring mechanism?}
\label{sec:classify-screw}

There are two broad classes of mechanism~\cite{LR-Landau} 
that can implement chirality on the microscopic (cell) level.
First, a {\it screw mechanism} is one depending on
conversion between longitudinal and rotational motion,
due to the inherent helicity of 
an actin filament or a microtubule.  
(``motion'' usually refers to a translocating motor, but
could instead be the fiber's growth, e.g. the 
microtubule screw-ratchet mechanism (Sec.~\ref{sec:mt-screw-ratchet})
and/or mere elastic displacement of its tip, e.g. in actin polymerization
(Sec.~\ref{sec:actin-pitch}).

Some screw mechanisms -- all of them conjectural -- 
   \begin{itemize}
\item[(a)]
actomyosin chirality
arising from the aziumuthal component of myosin power strokes 
(Sec.~\ref{sec:actomyosin-screw});
\item[(b)]
a helical Brownian ratchet in microtubule collisions 
(Sec.~\ref{sec:mt-screw-ratchet});
\item[(c)]
the change in pitch of actin fibers due to a change 
in their longitudinal load (Sec.~\ref{sec:actin-pitch}).
\item[(d)]
a pedagogical example (conceivably related to {\it Drosophila},
Sec.~\ref{sec:drosophila})
whereby a processive motor
that moves on its fiber with a helical bias (e.g. Myosin V) 
in the steady state tends to pull a cargo on the 
``right hand side of the road'',
see Ref.~\cite{LR-Landau}, Sec. 3.1.
\item[(e)]
the non-biological conversion mechanism 
in Ref.~\cite{gong-mt-ring},
whereby microtubules sliding on a kinesin-coated surface
tend to bend to the left. 
   \end{itemize}

The second route to implement chirality 
is that a specific linking molecule 
binds to the fibers with a rigid orientation.
and is anchored in the membrane, hence I 
call it an  ``anchoring'' mechanism.

Some examples of anchoring mechanisms are
   \begin{itemize}
\item[(a)]
the branching of microtubules in plant cell 
(Sec.~\ref{sec:plant-cell}, below);
\item[(b)]
the correct growth of a cilium out of a cell, in the
nodal cilia mechanism for vertebrates;
\item[(c)]
the conjectured kinked-attachment mechanism inducing a fiber to spiral
with a particular pitch inside a cylindrical cell
(Sec.~\ref{sec:kinked-attachment}).
%% \item[(d)]
%% and a vague alternative mechanism for actomyosin chirality
%% imagined in Ref.~\cite{LR-Landau}, Sec. 4.2.
   \end{itemize}
If a membrane-anchoring mechanism is involved, this furnishes a hint 
to biologists who hunt for pertinent genes using genetic engineering
or by screening for aberrant phenotypes following mutagenesis.
%%%%%%%%%%%%%%%%%%%%%%%%%%%%%%%%%%%%%%%%%%

Screw mechanisms, as they depend directly on helical symmetry,
appear more elegant in the context of physics.
By contrast, the anchoring mechanisms appear more natural 
in the context of molecular biology.
Underlying the screw/anchor dichotomy is one between 
two kinds of chirality:
(1) {\it absolute} -- the direct fixing of a third axis,
which realizes the absolute kind of chirality;
(2) {\it relative} -- the sign of a rotation relating two directions, 
which is realized by the screw motion.
Indeed, the screw
situation preserves rotational symmetry around the axis, as well
as translation along it;  axes are fixed indirectly, 
perhaps at the intercellular scale.
The respective collective dynamic states described here,
for both actomyosin in animal cells and for cortical microtubule 
arrays in plant cells, realize ``relative'' chirality
on a larger scale.
In either case, up/down asymmetry 
(with respect to the membrane normal) becomes converted to 
a chiral asymmetry within the layer.

A related but distinct dichotomy is that 
chirality may be implemented either through static 
positioning in the system, or else in its dynamics.  
Thus the attachment type scenario, realized
when the interacting molecules have a unique bonding site,
is not just absolute but static; by contrast, in the
screw scenario the binding is to a polymer with many 
symmetry-equivalent binding sites, so this is 
not only relative but dynamic.  

It will be seen
that a third combination is possible in principle, relative and static.
Such a situation is realized when parallel, microscopically helical 
fibers aggregate in an array with a cross-section that is 
(approximately) a triangular lattice.
As analyzed in Ref.~\cite{grason-bundles}, adjacent fibers ``want'' to be
rotated slightly about the line connecting them -- this is the same
generic asymmetry responsible for cholesteric liquid crystals.
Hence it turns the fibers form bundles of a parameter-dependent size,
each of which is a sort of twisted rope.  The continuum elasticity 
of this complex liquid crystal phase will be chiral too 
(in the ``relative'' sense). If cells were filled with such domains, 
their passive mechanical chirality could be {\it part} of a 
mechanism to implement chirality at the inter-cellular level.
However, the only biologically pertinent example I know of such an 
array is the coiling of viral DNA inside phase capsids~\footnote{
A ``phage capsid'' is the protein container of a virus that infects bacteria}
%%%%%%%%%%%%%%%%%%%%%%%%%%%%%%%%%%%%%%%%%

\subsection{Spontaneous symmetry breaking?}
\label{sec:SSB}

Another dichotomy in the scenario is whether or not 
a robust {\it spontaneous} symmetry breaking underlies
the chirality. Here, ``symmetry breaking'' refers that
a definite orthogonal axis is produced in \eqr{eq:rhrule},
but with a random sense, so that left- and right-handed
individuals are equally likely.
The systematic handedness is then controlled by an 
{\it indepedent} mechanism that supplies a {\it small} chiral biasing field.
This is analogous, in statistical mechanics, to the small magnetic field 
which determines the magnetization sense 
of an Ising magnet quenched into its {\it ferromagnetic} phase.

The alternative to spontaneous symmetry breaking
may be called a ``linear response'' scenario.
In this case, our system is the analog of the {\it paramagnetic} state
in statistical mechanics.
The left-right asymmetry is a smooth function of the chiral 
bias strength, and linearly proportional to it if it is small.
It follows that wild-type chiral bias 
must be larger in the ``linear response'' scenario than 
in the symmetry-breaking one.  
The ``linear response'' regime can be recognized biologically because
a a mutant missing the chiral bias remains symmetric and does not 
form a third axis in either direction.   
%%%%%%%%%%%%%%
\OMIT{Note, however, that when vertebrates
fail to realize the third axis, it seems to be a mutation
of downstream mechanisms that propagate the initial asymmetry
throughout the embryo (cite zebrafish)}

In the rest of the paper, except where noted, I implicitly assume 
the symmetry-breaking scenario.  This means that the sought-for 
mechanism need only produce a 
weak bias (let's arbitrarily say $\sim 10^{-2}$), 
since its effects get amplified hugely by
the (incipient) long range order.
A mutant missing the chiral bias forms a third axis 
just as well as the wild type, except the sign of $\yaxis$ is randomized.  
The latter phenomenon is
seen in human beings with Kartagener's syndrome (missing a
certain variety of the molecular motor dynein): 50\% of them
have a mirror-reversal of all internal organs, relative to
normal humans~\cite{mcmanus-RHLH}.

The mystery is why a chiral bias is needed at all, in the 
symmetry broken cases.  A mirror-reversed indidivual is
just as viable, and interactions between individuals only
occasionally depend on their having the same handedness.
Statistical physics can illuminate one
plausible evolutionary motive for the chiral bias.
Whenever L/R asymmetry is developed after an embryo is extended and
multicellular, spontaneous symmetry breaking cannot happen
instantaneously but must evolve over some time, during which
larger and larger parts of the system ``agree'' as to the final
sign.  This is analogous in statistical mechanics to {\it quenching} 
from a high-temperature disordered phase, into an ordered state.
In the absence of a bias field, overly rapid quenching leads to a 
mixture of several domains: in some of these the symmetry gets broken 
one way, in others it gets broken the opposite way.
In an  animal embryo, that woulld produce a spatial mishmash of 
normal and inverted regions, called ``heterotaxia'' in medicine, 
and responsible for serious congenital defects such as  the 
duplication or absence of organs.  By using a symmetry-breaking field, 
the embryo can afford to ``quench'' more rapidly 
-- time is often at a premium in devlopment -- yet not risk heterotaxia.

The above reasoning applies most strongly for phyla in 
which the fate of cells is fixed late in development,
by some form of patterning.  In examples where the style 
of development is to fix fates early -- even in
the first three divisions -- this concern appears less critical.
That suggests one possible reason why molluscs 
(an early developing phylum) 
have repeatedly speciated with a flip of the
chirality, and why the nematode {\it C. elegans}
may tolerate a high incidence of reversed individuals
(Ref.~\cite{okumura-review}, p. 3505).

The ``symmetry breaking'' notion has been discovered
independently in the context of development without help from 
statistical mechanics.  In particular, this may be
the most important notion in Brown and Wolpert's
ground-breaking paper \cite{brown-wolpert}.
Also, Wood's ``choice execution'' (Ref.~\cite{wood-review}, p. 76) 
is approximately the same notion as spontaneous symmetry breaking.

\section{Cell division and actomyosin layers}
\label{sec:actomyosin}

Animal cells lack rigid cell walls, and hence can realize
handedness (like other kinds of organization)
using motility, changes in shape, or displacements
with respect to adjacent cells.  The actin component of
the cytoskeleton exerts forces in two ways:
\begin{itemize}
\item[(i)]
Bundles of myosin molecular motors form bridges between actin
filaments, walking along them and thus causing relative motions
of the filaments. The actin and myosin combined constitute an 
active medium (called actomyosin)
%%%  which has a potential for dynamic instabilities.
\item[(ii)]
Actin polymerization generates an average 
tip force via the ``Brownian ratchet'' 
mechanism ~\cite{brownian-ratchet}.
Such forces extend the pseudopods (projections of the cell body)
that enable the (crawling) motility of eukaryotic cells.
\end{itemize}
This section takes up the first (actomyosin) story; 
I will return to the second (polymerization)
kind of mechanism in Sec.~\ref{sec:actin-pitch},
speculatively.

In this section, I lay out for the first time 
a comprehensive picture of actomyosin handedness 
at three levels, starting with the intercellular level 
where I first summarize the observations of spiral cleavage in
snail embryos (Sec.~\ref{sec:actin-intro})
-- {\it C. elegans} embryos are roughly similar --
and then show they follow if cell division is associated
with a particular twist (Sec.~\ref{sec:mollusc-division-twist}).
%%%(Sections~\ref{sec:latent-spindle} and \ref{sec:adhesion-constraint}),
The next level is a two-dimensional continuum theory for 
the actomyosin layer in one cell (Sec.~\ref{sec:actomyosin-continuum}),
where I point out that the aforementioned twist and two
other striking chiral behaviors are all consistent
with the {\it same} handedness of the continuum model;
lastly I present a microscopic-level screw-mechanism scenario, 
depending on the myosin stepping motion along an actin fiber
(Sec.~\ref{sec:actomyosin-screw}), which predicts the correct
handedness at the continuum level.

\subsection{Introduction}
\label{sec:actin-intro}

The initiation of left/right asymmetry has been studied 
carefully in two kinds of invertebrate, which have a 
pattern of development such that a cell's fate (i.e.
what kind of cell its descendants become) 
is determined at the few-cell stage: 
molluscs (specifically snails)
%%% ~\cite{kuroda-snail-actin}
and {\it C. elegans}.
In both cases the mechanism involves ``actomyosin'',
the layer of actin filaments and myosin motors immediately
under the cell membrane.
I conjecture that the ``collective level'' mechanism is
a twist about the division axis during each cell division.

\begin{figure}
\includegraphics[width=1.0\linewidth]{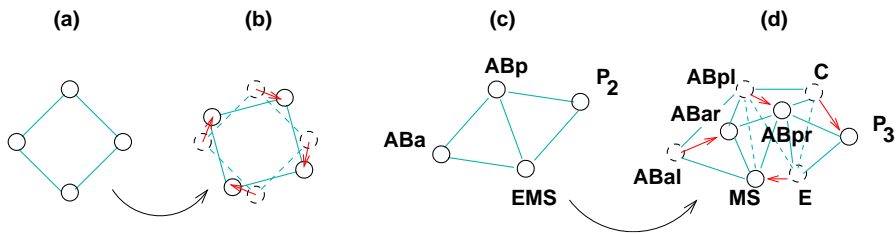}
\caption{Breaking L/R symmetry at the four-cell stage 
through division with a twist
(a,b) The 4-cell and 8-cell stages in the snail embryo.
Cell centers are represented schematically
by circles, and connected by lines if they adjoin (lower layer is dashed).
Arrows connect cells that divided.
From the nearly square embryo in (a), the four daughter cells 
are displaced around the 4-fold axis to the intercell furrows
as in (b), the ``spiral cleavage.''
(c,d) 4-cell and 8-cell stages in the asymmetric 
{\it C. elegans} embryo;
the displacement of dividing cells can be explained by 
a twist of the same sense as in the snail.  
Dorsal/ventral ($+z$/$-z$) and
anterior/posterior ($+x$/$-x$) axes are shown; the $+y$ axis
(right side) is out of the page.
}
\label{fig:celldiv}
\end{figure}

First I review some facts.
In snails~\cite{kuroda-snail-actin}
(specifically {\it Lymnaea stagnalis}),
after the first two divisions along mutually perpendicular axes, 
an embryo consists of four cells in a (nearly) perfect square.
%%%%%%%%%%%%%%%%%%%%%%%%%%
These divide vertically, producing four (smaller) daughter cells.
As is very long known, the square of daughter cells
breaks the symmetry by rotating $\sim 45^\circ$ {\it clockwise}
(as you look down on the square)
so as to sit in the crevices between the
original cells, the so-called ``spiral cleavage''
%%~\cite{kuroda-snail-actin}.
This is shown schematically in Figure~\ref{fig:celldiv}(a,b).
Which way this rotation goes decides the handedness of the final snail,
as proven by micromanipulation experiments in which 
the daughter cells were pushed the other way~\cite{kuroda-manipulate}.
It was earlier shown that a mutant which has
the opposite twist at the 8-cell stage also
ends up with an oppositely twisted shell~\cite{kuroda-snail-actin}.

In {\it C. elegans} embryos, the embryo with four
cells is also planar but unsymmetric.
The next cycle of divisions (to eight cells) makes it three-dimensional.
A cartoon of this process is shown in Fig.~\ref{fig:celldiv}, which 
compares molluscs and {\it C. Elegans}.
It can be seen that the screw sense of the motions is the same as
in the snail spiral cleavage, as first perceived
by Wood~\cite{wood-review,wood-review-2005}.
The handedness of the final embryo is determined by these divisions,
specifically at the six-cell stage as shown 
long ago by Wood using micromanipulations 
\cite{wood-review,wood-review-2005,wood-orig,poole-hobert}.
More recently Ref.~\cite{pohl-bao-Celegans} has 
carefully  studied all chiral asymmetries 
in the early divisions of {\it C. elegans}, 
including cell protrusions and adhesion.

%%%%%%%%%%%%% Supplementary parts for this section}  %%%%%%%%%%%%%%%%%

Right after the symmetry-breaking  division at the four-cell
stage, some chemical signal must be passed  between the cells,
depending on which cells are neighbors.
Since the neighbor relation has become asymmetric, 
this tells the cells which is L and
which is R; that ```bit'' of information is 
preserved in subsequent divisions for the descendents
of these cells, and is expressed functionally
at a much later stage.
The early-stage signal has been identified in the {\it C. elegans} case~\cite{poole-hobert}.

%%%%%%%%%%%%%%%%%%%%%%%%%%%%%%%%%%%%%%%%%%%%%%%%%%%%%%%%%%

Experimentally in snails, the twist depends on 
actin (it is disrupted by the depolymerization agent latrunculin)
but {\it not} on the microtubules~\cite{kuroda-snail-actin}
that (e.g.) constitute the mitotic spindle.
Incidentally, mutants that will have the wrong handedness 
initially stay {\it symmetric} at the $\to$ 8 cell division, 
but a {\it distinct} chiral mechanism (also involving actin) 
comes into play immediately afterwards; I will propose an
explanation in Sec.~\ref{sec:adhesion-constraint}.

\subsection{Intercellular level story for snails: twisted cell division}
\label{sec:mollusc-division-twist}

I start by addressing the ``intercellular'' level mechanism.
It is already known that the left/right asymmetry in snails is 
actin/myosin driven~\cite{kuroda-snail-actin}.

\subsubsection{Facts of cell division in snail embryo}
\label{sec:facts-snail}

Let us walk through the stages of the division with 
spiral cleavage.
Figure ~\ref{fig:spiral-axis}(a,b,c) show a cartoon of the
divisions in the wild-type snail embryo.
We are looking down from the animal ($+z$) axis.  The first division is
into cells called AB and CD. After the second division
[Fig.~\ref{fig:spiral-axis}(b)] the cells almost
form a planar square.

The third division axis is normal to the (approximate) plane of the
four cells, and produces four smaller daughter cells on the $+z$ side.
The defining event of ``spiral cleavage'' is that the daughter cells
rotate, clockwise (looking down the $z$ axis) in the ``dextral'' 
form, so as to lie in the furrows between the parent cells.
In the wild-type dextral, the spindle axis of the third division is
observed to be already tilted {\it before} that division (so called
``spiral deformation''~\cite{kuroda-snail-actin}).

The ``sinistral'' mutant is outlined in Fig.~\ref{fig:spiral-axis}((d,e,f).
In this case, there is no spiral deformation:
the daughter cells only rotate later on, during furrow ingression 
(i.e. the time they get pinched off from the mother cells) --
and they turn counter-clockwise (CCW), opposite to the wild type.

\subsubsection{Wild-type snail: latent-spindle mechanism}
\label{sec:latent-spindle}.

The observed (handed) positioning of the cells is compactly explained 
if we posit that two dividing cells acquire a relative twist,
always of the same screw sense
-- let's say {\it clockwise}, as you view one cell from its neighbor.
This ``{division twist}''
must be driven by flows of the actomyosin layer 
associated with the cell membrane, which (in particular)
concentrates along the dividing plane and forms the contractile
ring that pinches off the daughter cells.  
(The large-scale, chiral flow will be addressed 
in Subsec.~\ref{sec:actomyosin-continuum}
at the single-cell continuum theory level.)
Then, spiral cleavage can follow as a purely mechanical 
consequence.  But I find {\it two} alternative
mechanisms, derived from the {\it same} mechanics 
posited above but acting at {\it different} times 
relative to the division, that produce rotations 
of the four daughter with {\it opposite} senses.
Both mechanisms are illustrated in Fig.~\ref{fig:spiral-axis}.

\TEMP{So let's hypothesize a torque which
always drives the same twist of two 
daughter cells about their axis of division.  
If the cells were already latently polarized along parallel
axes for the subsequent division, those axes get
tilted by the torque as observed.}

We can call the first mechanism the ``latent spindle.''
The key assumption specific to 
this mechanism is that, even before one division is complete, 
the mitotic axis to be used for the {\it next} division (along a 
perpendicular axis) exists, in some latent form.  
The physical determinants of this might be centrioles 
(molecular complexes in the cell which, when present, anchor the
spindle's endpoints); being attached to membrane, they will get 
carried along by the large-scale actomyosin/membrane flow.
Indeed Wood (Ref.~\cite{wood-review}, pp. 74-75) specifically suggested 
centriole positioning as an ingredient of the mechanism
for snails, though in a different way than I proposed.  
(He suggested distinctive segregation of the inherited 
and the newly formed spindle poles that are at either 
end of the division axis.)

Initially, the latent axes are parallel in the two cells;
but the latent axes get carried along by the division twist
of the prior division as it proceeds, so they get tilted 
relative to each other.
Then, as figure \ref{fig:spiral-axis}(b) shows,
the combined twists from the first two divisions imply spindle 
alignments for the third division that are predicted to
be tilted around the fourfold axis exactly as seen experimentally
in the four-cell stage, 
called the ``spiral deformation''~\cite{kuroda-snail-actin}.
When the daughter cells subsequently divide along these tilted axes,
they are biased with a {\it clockwise} rotations relative to the mothers,
as seen experimentally in wild type
[Fig.~\ref{fig:spiral-axis}(c)].

\subsubsection{Mutant snail: adhesion-constraint mechanism}
\label{sec:adhesion-constraint}.

The second mechanism might be called ``adhesion constraint.''
It means a tendency to twist about each division axis that 
is frustrated by the constraints of adhering neighbors, but 
drives twist by larger groups of adhering cells.
Start again with the square of cells resulting from the first 
two divisions; imagine that the four daughter cells, before 
separating from the four large mother cells, formed adhesions with
each other.  What does the {\it same} postulate predicts
for the dynamics {\it after} (or while) the small daughter cells 
divide?

\begin{figure}
\includegraphics[width=0.88\linewidth]{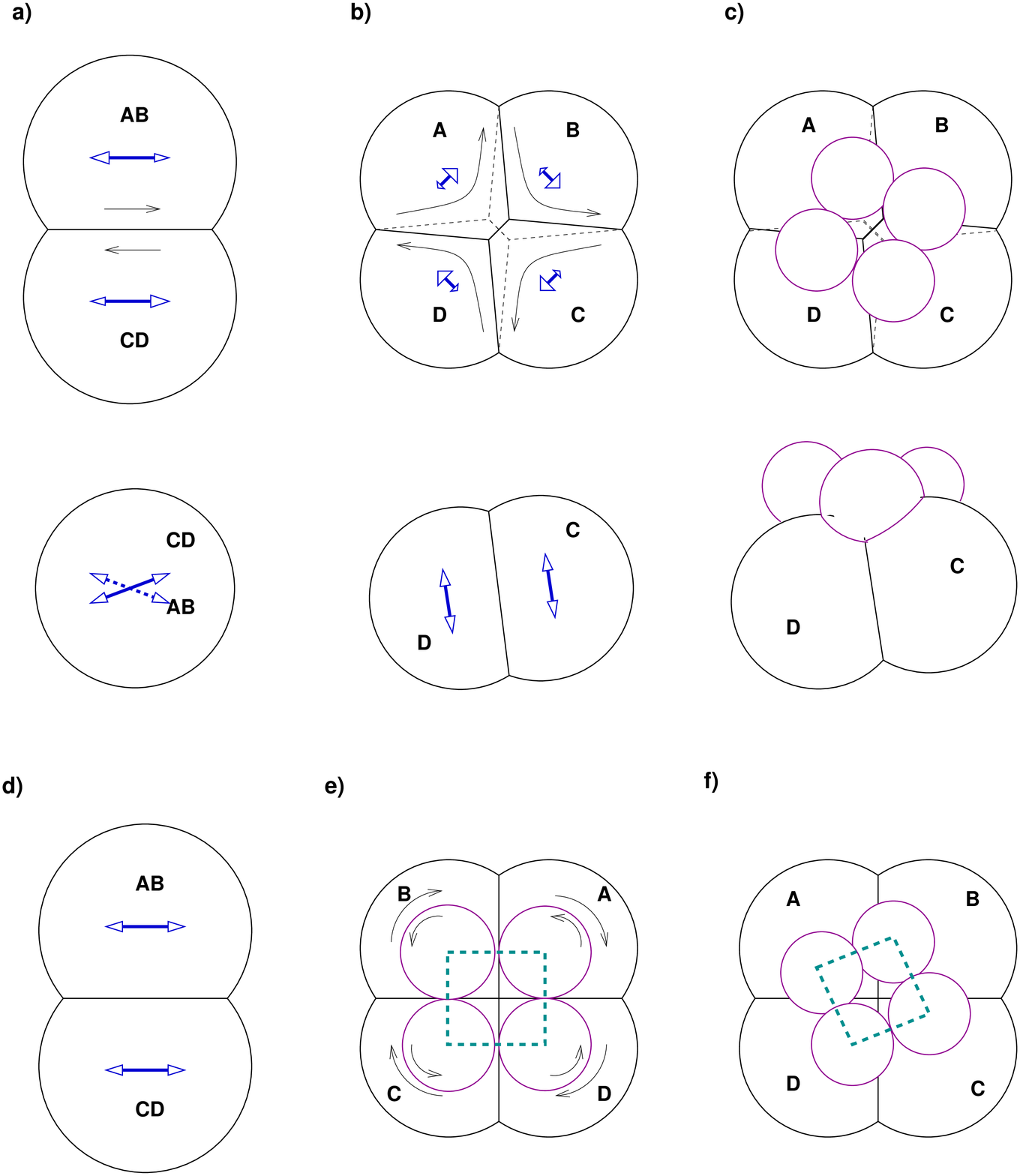}
\caption{
(a)--(c) Early divisions in wild type (sinistral) snail,
two views of each (left: down $z$ axis, i.e. from
the ``animal'' direction; right, down $x$ axis).
Thin arrows show the predicted large-scale cortical flow;
heavy arrows show a latent (incipient) spindle direction.
(a). two-cell stage (named AB and CD)
(b). four-cell stage; the next axis is almost normal
to the plane of the cells
(c). eight-cell stage.
Parts (d)--(f) are a cartoon of divisions in the dextral mutant snail
(d) The latent spindle directions are not twisted in the first
division (nor in later ones) (e) After the four daughter cells
bud off, the actomyosin layer flows as indicated by the arrows.
Since the four daughter cells are already connected by
cell adhesions (symbolized by dashed square) the combined effect 
of the flows is for them to rotate as a rigid body, in the
sense shown in (f).
Note that, relative to the cell boundaries, the actomyosin
flow has the {\it same} sense in (b) and in (e), but this
leads to {\it opposite} outcomes in (c) and in (f), depending
on whether that flow is active before or after division.}
\label{fig:spiral-axis}
\end{figure}

If we had isolated mother/daughter pairs, 
torques due to the cortical flows would drive 
each daughter cell to rotate {\it counter-clockwise} 
around the division axis  (looking down the $z$ axis),
as shown by the arrows in Figure~\ref{fig:spiral-axis}(e).
But after  the four daughter cells form adhesions, they
are no longer free to rotate separately, but only
as as a rigid assembly.  The net torque on this 
combination turns the square of four daughter cells 
counterclockwise, i.e. {\it rightwards} as seen from
the mother cells, which is {\it opposite} to the sense
of rotation of the latent-spindle mechanism of 
Sec.~\ref{sec:latent-spindle}.

Thus, the {\it same}
mechanism produces {\it contrary} effects, depending on
the time in the division process at which it acts.
This suggests the hypothesis that, in the snail {\it L. stagnalis},
both kinds of handedness can be explained by a single
chiral mechanism -- the difference is ascribed to a non-chiral
property; namely, it appears that the spindle-orientation
effect is the stronger effect in the (sinistral) wild type,
but is missing in the (dextral) mutant, so that the secondary
division effect takes over.

\subsubsection{Tests of division-twist idea?}

An immediate consequence of the idea is visible
already at the four-cell stage. 
If the first divisions have the {\it same} sense of twist 
as needed to predict the spiral cleavage,
the embryo should be slightly warped, with the A and C cells
raised (so they touch on the $+z$ side) and the B and D cells
lowered (so they touch on the $-z$ side), as illustrated
in Fig.~\ref{fig:spiral-axis} (b).
This agrees with the sense of warping observed 
experimentally (Ref.~\cite{kuroda-snail-actin},
figure 2B).

On the other hand, the mutant sinistral {\it L. stagnalis} 
should lack the main warping; indeed, according to the
adhesion-constraint scenario, in the second division
cells B and C should twist as a unit 
after they separate from cells A and D so the warping
should occur in the reversed direction. Indeed, 
figure 2B of Ref.~\cite{kuroda-snail-actin} shows 
an absent or perhaps slightly reversed warping
in the mutant.

Of course, the last observation only shows that the twist
occurs in different cell divisions.  To test my proposal that
the {\it same twist sense} acting at different times can 
produce the observed effects, one apparently needs mutations
or drugs  that change the duration of the respective stages,
or that change the strength of cell-cell adhesion governing '
the adhesion-constraint mechanism.  A mechanical test might
be to remove three of the four daughter cells from the mutant,
in which case the remaining cell (freed from contraint)
ought to rotate around the axis joining it to the parent cell.

\OMIT{
Geometrically, this object
approximately has the point symmetry group called ``$222$'' (or ``$D_2$''),
with three mutually perpendicular twofold axes.
However, we must take into account the preexisting distinction of $+z$
versus $-z$ directions (animal versus vegetal axes) as well as the 
inequivalence of the first division plane to the second. 
Thus the exact (functional) symmetry group
is only ``$2$'' (or ``$C_2$''), having one twofold axis aligned with $z$.}

\subsection{Micro story: continuum equations}
\label{sec:actomyosin-continuum}

I suggested in the last subsection that 
the ultimate handedness both snails and {\it C. elegans}
would be explained if the actomyosin
layer, around the division furrow where it concentrates,
shears in a {\it clockwise} (CW) sense, as you look down at the
membrane, as marked by the small arrows in Fig.~\ref{fig:spiral-axis}
and by the straight arrows in Fig.~\ref{fig:somersault-sense}(c).
motion in the actomyosin layer, around the time of cell
division, can drive a twist between the two dividing cells.
In fact we can see two other striking chiral phenomena 
{\it with the same clockwise sense} in actomyosin cortical layers
(subsec.~\ref{sec:actomyosin-layers-sense}); this would be
explained if pairs of the units forming the layer interacted
with a clockwise twist, which can be represented at the collective
level by a single ``skew stress'' parameter 
(subsec.~\ref{sec:actomyosin-continuum}).

\subsubsection{Chiral phenomena in actomyosin layers}
\label{sec:actomyosin-layers-sense}

Animal cells typically have a cortical (adjacent to membrane)
actin-myosin layer.  This practically two-dimensional medium 
is responsible (in some theories) for the 
effective surface tension of cells~\cite{manning-tension}.
In the form of the ``contractile ring''~%
\cite{contractile-ring,kamasaki-contractile,prost-contractile,astrom-contractile},
%%% vavylonis-contractile,
%%% gov-contractile})
an array of roughly parallel filaments that contract to pinch off
the two cells from each other.
Note the myosin in these layers is a form of myosin II, the same
myosin found in muscles, which is a non-processive motor protein
(does not stay on the actin for many steps).

Massively chiral behavior is already known in the
actomyosin layer in two instances, both in egg cells
(convenient for imaging since they are so much larger than 
regular cells).

\begin{figure}
\includegraphics[width=0.75\linewidth]{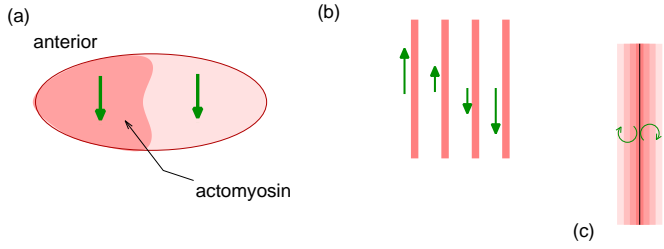}
\caption{
Observed chiral phenomena in actomyosin layers.
(a) Egg cell of {\it C. elegans} at end of the polarization in
which actin (shading [pink online]) concentrates towards the 
anterior end.  Just before the division furrow forms, a concerted
motion is seen as indicated by the large arrows.
(b) Schematic of the {\it Xenopus} egg, in which bundles of 
actin (stripes) around the incipient contractile ring
were observed in Ref.~\cite{danilchik-xenopus-twist}
to shear with relative velocites as shown by the arrows.
(c) A band of actin is shown with densities 
represented by shading [pink online].
The continuum theory of Sec.~\ref{sec:actomyosin-continuum})
with $q>0$, or equivalently a clockwise interaction of pairs
(small arrowed arcs), gives no net force 
in a uniform distribution of actin, or 
at the density maximum in the middle, but
tends to drive a clockwise  shear on either side,
in accord with the motions assumed in 
Figure~\ref{fig:spiral-axis}.}
\label{fig:somersault-sense}
\end{figure}

\begin{itemize}
\item [(1)]
The {\it C. elegans} egg, before its first division,
develops a polarity which includes a segregation of the actomyosin
layer towards the anterior end, shown by a denser shading
in Fig.~\ref{fig:somersault-sense}(a).
Towards the end of that process, all the actomyosin rotates 
around the anterior-posterior axis~\cite{Celegans-somersault},
in the {\it clockwise} sense, as one faces the anterior end
from outside the cell.
%%% {I checked movie, it is CW.}
(This phenomenon is familiar to {\it C. elegans}
researchers and visible in movies, but has not been commented 
on in print to my knowledge.)

\item [(2)]
Danilchik {\it et al}~\cite{danilchik-xenopus-twist}.
treated {\it Xenopus} (frog) eggs 
with a poison (butanedione monoxime = BDM) that partially disrupts the actin network.
When the contractile ring begins to form along an egg's future cleavage plane,
it develops an array of parallel actin bundles shearing each relative to the next, 
such that the local vorticity is {\it clockwise}, as
abstracted in Fig.~\ref{fig:somersault-sense}(b).
This mechanism does
{\it not} depend on actin polymerization (as checked by the null effect
of latrunculin, which inhibits polymerization).
\item [(3)]
Finally, of course, there are the early divisions in snails
and in {\it C. elegans}, both implicating actin as detailed
in Sec.~\ref{sec:actin-intro}; as illustrated in Fig.~\ref{fig:spiral-axis},
this also entails a clockwise shear around the divsion furrow, as
cartooned in Fig.~\ref{fig:somersault-sense}(c).
\end{itemize}

\subsubsection{Form of continuum equations}
\label{sec:actomyosin-continuum-form}

Our philosophy is to seek the simplest set of
continuum equations that are compatible with symmetry and Newton's
law of action/reaction, and that have a chiral term.
(Note that our layer is in contact with the membrane above
and the cell fluid below, which exert forces such as 
viscous damping, so our equations need not be invariant under boosts.)

%% The basic idea is inspired by one conjectured microscopic mechanism 
%% (Sec.~\ref{sec:actomyosin-screw})
%% but a larger class of mechanisms give the same continuum result.
The necessary ingredient is that the ``units'' constituting our
medium have a  pairwise 
interaction that tends to rotate them in a fixed sense
(say clockwise) about the midpoint between them.   
Naively, one might expect this to drive the velocity field
to have a vorticity, 
$\partial v_y/\partial v_x- \partial v_x/\partial v_y < 0$.  
However, so long as the density is
uniform, the mean force on a given ``particle'' is zero by symmetry, 
since it has (on average) equally many nearby particles on either side.
Therefore, chiral forces are associated with {\it gradients}
in the actomyosin density.

I believe the simplest functional form for the continuum theory
of a two-dimensional actomyosin layer is that the stress tensor 
$\sigma(\rr)$ acquires the form
   \beq
  \ssigma(\rr)  \equiv \pmatrix
           {\sigma_{xx} & \sigma_{xy} \cr
           \sigma_{yx} & \sigma_{yy}} =
  \pmatrix{ p   &  q \cr
           -q   &  p } + \ssigma_{drag}(\rr)
   \label{eq:stress-T}
   \eeq
where $p$ is the (two-dimensional) pressure,
$q$ is an unusual ``skew stress''
($q>0$ for the CW interaction that I posited);
$\ssigma_{drag}$ is the contribution of drag to the stress tensor.
Both $p(\rr)$ and $q(\rr)$ should be functions of the local number density 
$\rho(\rr)$ (and any other collective state variables)
according to some constitutive equation.
To {\it rigorously} define $q$ (or even $p$), 
one would need for the actomyosin to form a statistically uniform and
stationary steady state.

It is a general dictum in mechanics that $\ssigma$ should
be a symmetric tensor, but that is not logically necessary when 
torques can be applied in the bulk interior, as in many
actively driven systems.  In the case considered by
Ref.~\cite{Fu11}, the torque is due to rotations
by each particle around its center of mass; in our
case, it comes from the interaction of two particles 
coupled to a membrane and cell fluid that define a preferred frame.
Specifically, the component $\sigma_{yx}$
(e.g.) is defined as the $y$ component of force being exerted on
the $x<0$ side of a plane normal to the $x$ direction.
Clearly, whenever a unit-unit interaction crosses that plane, it contributes
a force in the $+y$ direction, meaning $\sigma_{yx}=-q<0$,
similarly $\sigma_{yx}=+q >0$.

Often these systems have another state variable besides the
density, namely the bidirectional (or ``nematic'' in statistical
physics) orientational alignment of filaments.  In that case,
the stress tensor has an additional traceless symmetric term 
with its axes aligned with the nematic axes. Furthermore, 
in our chiral systems the stress axes will generically get
rotated a bit from the nematic axes (in a definite sense).  
I will not consider this complication in the present paper.

The consequence of \eqr{eq:stress-T}
is that the local force density (per unit area) is
     \beq
       \FF(\rr) = \nabla p + \hat{\zz} \times  \nabla q   -\lambda \vv
            + \mu \nabla^2 \vv
       \label{eq:F-Dq}
     \eeq
where $\hat{z}$ is the unit normal to the layer. 
%%%%%%%%%%%%%%%%%%
The last two terms are, respectively, drag due to the membrane,
and drag due to shear of the actomyosin layer.
The first drag term models the viscous damping as the actomyosin slides
past the (fixed) membrane and 
the bulk fluid (cytosol) of the cell 
which is entrained by the cytoskeleton that permeates it.
Inertia is negligible at our low Reynolds numbers
so the velocity field $\vv(\rr)$ is the solution of $\FF(\rr)=0$,
%%%%%%%%%%%%%%%%

\subsubsection{Dresden three-dimensional continuum equations}

The MPIPKS (Dresden) group~\cite{Fu11,strempel}
%%%F\"urthauer, Grill, and J\"ulicher 
have been developing a comprehensive three-dimensional
continuum description of active, chiral media,
which is less ad-hoc than the one presented here,
but also more complicated.

The key differences are  that 
\begin{itemize}
\item[(i)]
They model a {\it three} dimensional actomyosin fluid.
Presumably an effective two-dimensional model can be derived 
from it, so as to model the (relatively thick)
cortical actomyosin layer by a density 
that falls off with distance from the membrane.  
\item[(ii)]
Their model takes into account not only the 
angular momentum inherent in the velocity field, but separately
the angular momentum due to internal rotations of the microscopic 
rigid units constituting the active fluid.  
This seems motivated by the case that the active entities are
swimmers. In a sufficiently damped limit, as in the case of
the actomyosin surface layer,
one can eliminate this kind of orientational degree 
of freedom and obtain an effective model in which 
angular momentum injected by the active degrees
of freedom is attributed directly to the center-of-mass
degrees of freedom.  
\end{itemize}
Overall, their philosophy is to incorporate all possible terms 
allowed by symmetry, which means quite a few unknown 
coefficients.  Here, in contrast, I tried to adopt the 
{\it minimal} model that is internally consistent and appropriate to
our particular system. 

\subsubsection{Applying the equations}

The motivation for this continuum theory was the cell
division story of Sec.~\ref{sec:mollusc-division-twist}.
What does the second term in \eqr{eq:F-Dq} say about
dividing cells?
%%%%%%%%%%%%%%%%%%%%%%%%%%%%%%%%%%
The contractile ring and furrow, along a
dividing cell contact line (as seen in Fig~\ref{fig:spiral-axis}),
is a local maximum of the actomyosin density.
In light of the density gradient in either direction
away from the line,
the chiral dynamics of \eqr{eq:F-Dq} imply a clockwise shear adjacent to 
these boundaries; the arrows shown in Fig.~\ref{fig:spiral-axis}
correspond to $q<0$ in \eqr{eq:F-Dq}, as shown in
Fig.~\ref{fig:somersault-sense}(c).

The same equations could be applied to a different example of
emergent chirality in which the elementary units are cells 
rather than actin fragments, and the pair interactions are due to 
actin-driven cell motility and/or cell-cell adhesion, rather 
than to myosin bridges.  This is applied in Sec.~\ref{sec:mammalian-channels}
to spontaneous chiral flows in a population  of eukaryote cells.

The anterior/posterior polarization  in the {\it C. elegans} egg can be
represented by an exernally imposed gradient in the 
actomyosin activity rates, 
analogous to the enhanced activity postulated in a band which 
triggers formation of the contractile ring~\cite{astrom-contractile}.
%%%%%%%%%%%%%%%%%%%%%%%%%%%%%%
The rather uniform rotation of the entire actomyosin layer
does not follow easily from the Eq.~\eqr{eq:F-Dq}, which 
describes a flat layer and assumes the total {\it force} 
in that layer is zero.  But we fit this into the
same framwork if we step back from the continuum theory 
to the underlying idea of a density-dependent clockwise torque.
The anterior end contributes a net torque around the long
axis in the CW sense (as seen from that end), while the 
posterior end contributes with the opposite sense;
but the anterior term is much larger (as the layer is
denser there). Thus if all parts of the actomyosin layer
are rigidly linked, the whole layer rotates in the sense
of the combined torque, which is CW as observed.

In principle, many experimental tests are possible using
video data of the actomyosin layer as already used in 
Ref.~\cite{mayer-Celegans}.  Eq.~\eqr{eq:F-Dq} implies
visible local velocities transverse to any variations in the 
density, whether they be random statistical fluctuations,
or cuts in the cortical layer made by 
laser ablation~\cite{mayer-Celegans}.

\subsection{Actomyosin: molecular level story}
\label{sec:actomyosin-screw}

What microscopic mechanism can implement the chiral inter-unit force?
By the fundamental symmetry notion invoked in 
Sec.~\ref{sec:question-assumptions},
any such mechanism must involve the membrane. 
(In its absence, the system has a symmetry of swapping the up and down 
directions by a 180$^\circ$ rotation, but the handedness would be
opposite when viewed from the down side.)  

I will develop an explanation of the ``screw'' class
(in the categories of Sec.~\ref{sec:classify-screw}),
i.e. depending on the actin fiber itself being a helix,
and using  myosin II (the kind in the actomyosin layer
as well as in muscle cells).
To start, note that in a standard motor protein assay,
in which actin filaments glide on a surface coated with myosin II, 
actin is found to twirl CW looking towards 
the plus end~\cite{myosin-twirling}.
(CCW in the direction of filament motion; it has not been 
explained why an opposite sense was reported in the 
past~\cite{nishizaka-twirling}.)
This indicates a CCW azimuthal component of myosin II's step 
(as it moves towards the plus end).

The detailed theoretical explanation~\cite{vilfan-twirling} shows
that the rotation depends on a commensurability of the
myosin step with the lattice of binding sites on actin
(considered as a two-dimensional lattice wrapped around a cylinder).
Thus the handedness of the typical step could alternate in sign 
as the myosin length is monotonically reduced (as is engineered in some
experiments).

\subsubsection{Details of myosin up/down mechanism}

The dynamics of actomyosin depend on
myosin bridges that connect two actin filaments, with 
multiple myosin heads at either end~\cite{hyman-PAR-rev}.
Due to the multiple heads, the bridge's motion along the
filaments might effectively be processive, even though
individual myosin II molecules are not processive.
(Previously, the processive myosin V had been observed to spiral 
as it traveled along an actin filament~\cite{ali-helical}.)
This partially justifies our drawing the myosin as if
it stayed processively on the actin, but that picture is
only for convenience: if the myosin only gives one power
stroke before releasing, it will still drive a net twisting
of the actin if that stroke has the azimuthal component.

In contrast to (say) the thin plant cell microtubule layer
of Sec.~\ref{sec:mt-plants}, the actin cortical layer 
in animal cells is 100 or even 1000 nm thick~\cite{morone-thick}.
This contains many actin fibers extending transverse to the
membrane (from which they nucleate and grow in various
directions into the cell fluid). This offers an opportunity
to relate the handed motion of the myosin II 
around a filament, to the handed motion orbiting that we posited
in subsec. ~\ref{sec:actomyosin-continuum-form}
between two of the units (namely two actin filaments).
Note that the actin nucleates on the membrane via formins
which are associated with the plus end, so actin filaments
are polarized {\it towards} the membrane (``upwards'').

\begin{figure}
\includegraphics[width=0.57\linewidth]{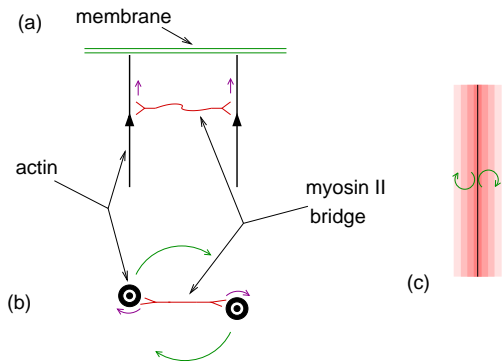}
\caption{Handed forces due to azimuthal motion of myosin II
(a). Side view of two actin filaments, polarized upwards as 
marked by large arrowheads on the filaments, and connected by
a myosin bridge which walks upwards as marked by small
arrows [purple online].  Only one myosin is shown at
either end of the bridge; actually there are many.
(b).  Top view (from outside membrane) of the same
filaments.  As the myosin II motors step towards us,
the screw component of their motion is in the direction shown
by the small arrows [purple online].
This tends to ``roll'' each filament CCW, which in turn
``rolls'' against the fluid to drive the pair in a clockwise
motion (large arrow arc [green online]). 
(c). same as Figure~\ref{fig:somersault-sense}(c).
A band of actin is shown with densities 
represented by shading [pink online].
The clockwise relative interaction of pairs (arrow arcs [green online]),
gives a cancelling mean force 
at a symmetry center such as the band's center.
But in the band's wings where the density is different 
on either side, there is a net force which follows a
clockwise pattern on a larger scale.
}
\label{fig:vertical-bridge-hand}
\end{figure}

Consider two parallel actin filaments, both polarized
upwards,  and connected by a myosin bridge,
as shown in Figure~\ref{fig:vertical-bridge-hand} (a,b).
Each myosin head (or multi-head) walks upwards, tending to 
spiral CW around the actin (as viewed from above the
membrane, opposite to their direction of motion).
However each is constrained by the bridge connecting
it to the myosin head on the other filament.
This produces a net torque which tends to rotate the
pair CW around the middle of the bridge, the very 
clockwise force we had posited.

\subsubsection{Questions about this theory, and experimental tests}

It is not easy to graft this handedness notion
onto a quantitative model of the actomyosin layer, for
no such model exists: too few of the parameters
have been measured experimentally in actomyosin.
The existing models for actin contractility~\cite{Va08,Zi08} 
are qualitative at heart.

The microscopic screw-mechanism hypothesis, presented here,
implies a universal sense of handedness due to the 
intrinsic properties of actin and myosin II.  
If we take this seriously, it makes a powerful prediction:
reversed handedness is possible, but its mechanism can 
{\it never} be functionally equivalent to the wild-type's
at the bottom level.  There is one qualification:
as handedness is propagated from its
original cause, either from lower to higher levels in these
mechanisms of the origin, or from upstream to downstream 
in the signaling cascades that specify every cell's fate
in development, the response in the reversed organism 
probably does become functionally symmetry-reversed 
at some point.  In snails, though, the external shell
shape is subtly different between 
the mutants and wild type~\cite{snail-shape-dist};
of course, this may be one of the many independent effects
expected from a mutation affecting the cytoskeleton.

At first glance this sweeping prediction seems to be perfectly 
confirmed in the qualitative difference between dextral and sinistral 
mutant snails observed in Ref.~\cite{kuroda-snail-actin},
as summarized in Sec.~\ref{sec:adhesion-constraint}.
the dextral snails.
However, a further observation seems to contradict it.
Namely, snails that are sinistral {\it as a species} show a 
spindle-orientation effect (spiral deformation) in the
opposite sense.~\cite{kuroda-snail-actin}. 
Furthermore, the difference in external shell shape 
between wild-type and mutants also gets flipped in those species 
~\cite{snail-shape-dist}.
We expect the chiral sense of the large-scale flow to 
depend on universal properties of non-muscle myosin II, 
and hence not to flip its sign from species to species.
If further observation really confirms that the
sinistral species are perfect mirror images, 
we might be forced to abandon the screw mechanism
for the ``anchoring'' type mechanism
(in the classification of Sec.~\ref{sec:classify-screw}).
The respective anchoring molecules would not 
be mirror-images of each other; all that is necessary
is that they would bind myosin molecules with the
heads oriented in opposite directions.

Of course, to flesh out the physical picture presented above,
it ought to be extended at least to give an (order-of-magnitude)
estimate of the expected twist rates.  
\OMIT{o check whether the observed asymmetries can develop
in the available time.}
This is not hopeless:
using hydrodynamics we can certainly estimate the chiral component
of the force between two vertical filaments from known parameters
of the motors. The main unknown is the number of such myosin
bridges, which -- conceivably -- might be related to the 
contractile force within the layer as measured in 
laser ablation experiments~\cite{mayer-Celegans}).

%%%%%%%%%%%%%%%%%%%%%%%%%%%%%%%%%%%%%%%%%
\section{Plants and 2D microtubule array}
\label{sec:mt-plants}

The origin of handedness in plants
involves another dynamic, cortical fiber array
-- not of actomyosin as in animal cells (Sec.~\ref{sec:actomyosin})
but of microtubules (mt)~\cite{hashimoto-mt-review}.
A major difference is
that whereas few physical parameters are known for 
the actomyosin array, they are all quantitatively
measured for the plant cell microtubule arrays.
The microtubule array array is a template for the
cellulose fibers formed outside the membrane, which 
forms the cells walls characteristic of being a plant.

\subsection{Macroscopic root behavior}
\label{sec:macro-root}

On the large scale, many species exhibit
systematic handedness in their roots, shoots,
or flower shapes:
\begin{itemize}
\item[(1)] 
The growing root's tip tends to trace out a helix
with a consistent sense
(flagrantly so in climbing vines).
\item[(2)]
Roots growing against a tilted hard surface have a handed bias
(in that plane) from the most directly downward direction,
called ``root skewing''~\cite{hashimoto-I}.
\item[(3)]
The ``cell files'' -- longitudinal rows of cells on the
outside of the root -- twist around the root's axis.
\item[(4)] 
The cells (long cylinders aligned with the root's 
macroscopic axis) have membrane-associated arrays 
of parallel microtubules which {\it spiral} around
the cell in a characteristic sense.
\end{itemize}
In the model species {\it arabidopsis},
mutations that reverse the sign of one behavior 
also reverse the others~\cite{hashimoto-I,hashimoto-2007}, 
suggesting that the microtubule array is the microscopic origin.  
Indeed, mutants with  strongly helical behaviors  turned out to be
mutants of the tubulin molecule itself~\cite{hashimoto-2007}.

Can the macroscopic behavior, at least, be explained from mechanics 
and simple biological models of the root growth?
Recent experiments have elucidated root twist by three-dimensional imaging
of {\it Medicago Truncatula} roots, growing downwards in clear gel.
The cell files are untwisted until the tip encounters a resisting
layer, upon which they become helical (and the root's buckling develops
in a helical form) with a reproducible bias in handedness~\cite{jesse-roots}.

Thus, it appears the twist is an active response to sensing of a
{\it longitudinal} force by the tip, an unusual example of the
touch responses known as ``thigmotropisms''; the roots also have
an active response to gravity called ``gravitropism''.  These
tropisms usually are not chiral; they involve a transverse bending of
the root in a direction aligned with the transverse force, or the
component of gravity as sensed by the tip; it is transmitted
somehow by the hormone auxin to the region a short distance 
back from the tip where new cells are elongating, which implement
the bending by elongating more on one side than the other.
The observed phenomenon of sinusoidal-like alternate bends
(``root waving'') can be understood in such a scenario~\cite{waving-holbrook}.

The intermediate organism-level step, how the microtubule 
array's handedness implies the macroscopic chiral thigmotropism,
is still unclear.  Hashimoto~\cite{hashimoto-PC} has suggested
a passive mechanical explanation: the inner layers elongate
deficiently (the strongly biased mutants tend to have defects in
the microtubule array leading to loss of cylindrical cell shape),
while the epithelium elongates more, which in turn triggers
a helical buckling of the latter.  However, this picture in itself
is a spontaneous symmetry breaking:  some small biasing is
still needed to give it a regular handedness.

It can be noted here that an early mechanics paper
related the macroscopic helical growth  of the fungus 
{\it Phycomeces} to the spiral arrangement
of the chitin fibers forming its cell walls, analogous to the cellulose fibers
in plant cell walls~\cite{gamow}.  
The cytoskeleton in fungi is much less thoroughly
studied than in animals or plants; since they are more closely related
to animals than plants, one could speculate that an actin cortical
layer takes over the role of the microtubule array in plant cells.

\subsection{Cell level mechanism: microtubule arrays}
\label{sec:plant-cell}

I turn to the cell level mechanism.
Video microscopy shows that
microtubule arrays get oriented in a collective process 
governed by the following rules:~\cite{dixit-cyr-data,mt-rules}
\begin{itemize}
\item[(1a)] {\it Growth}:
The $+$ end of the mt grows quickly, 
while the $-$ end steadily depolymerizes,
more slowly  than the $+$ end grows.
\item[(1b)] {\it Spontaneous catastrophe:}
the $+$ end can spontaneously transition into a 
rapidly {\it depolymerizing} state,
so that the whole microtubule may soon be undone:
this is called a ``catastrophe''.   There is also
a rate for transitions back to the growing state.
\item[(2a)] {\it Collision and entrainment}:
When a growing mt hits another at a relative
angle less than $\sim 40^\circ$, it bends 
and {\it entrains} parallea; this process generates bundles..
\item[(2b)] {\it Collision-induced ``catastrophe''}:
If the angle is larger, the growing mt suffers a ``catastrophe''
\item[(3a,b)] {\it Nucleation and branching}
New mt's nucleate both (a) at random places and (b)
on existing mt's; the latter have branching angles centered around
$\sim 40^\circ$ or so from the $+$ direction~\cite{chan-mt-data},
but also a significant number are at angle 0, 
i.e.  are born entrained ~\cite{chan-mt-data}.
\end{itemize}
The rates for these processes have been measured 
experimentally~\cite{dixit-cyr-data,shaw-ehrhardt-data,kawamura-wasteneys-data,chan-mt-data}
and used recently in numerous mathematical models~%
\cite{baulin-mt-sim,hawkins-mt-math,tindemans-mt-sim,allard-mt-sim,%
shi-mt-sim,eren-mt-sim,deinum-mt-sim} which have been reviewed in
Ref.~\cite{eren-review}.
It still unsettled~\cite{eren-review} whether the parameters
in Rule (i) put the single-mt behavior just above or just
below the threshold at which it would grow without bound;
this question is not critical in the array state, where
mt growth is actually limited by collisions [Rule (iii)].
Processes (ii) and (iii) both tend to drive the mt's into
a steady state in which microtubules tend to be parallel
or antiparallel to some direction in the plane which is
(so far) arbitrary;  such a bidirectional, symmetry-broken 
phase is labeled ``nematic'' in statistical mechanics.
(Experimental arrays actually seem to be polarized in
{\it one} direction, not bidirectional~\cite{chan-mt-video},
but this point is unimportant for the chiral properties.)
%%% Another reference: 
%%% R. Dixit, E. Chang, and R. Cyr, Mol Biol Cell 17, 129801305 (2006),
%%% "Establishment of polarity during organization of the
%%% acentrosomal plant cortical microtubule array"

A less clear aspect of the array alignment story is, why does
the observed array orientation have a systematic relation  --
mostly transverse, but tilted chirally-- to the cell's long axis?
The widely accepted explanation for this is the cell caps~\cite{ambrose-CLASP},
a sort of finite size effect.
One can treat the cell's entire membrane as roughly a flat rectangle 
having periodic boundary conditions in the azimuthal direction, 
but open boundary conditions the other way, representing the cell's end caps.  
These boundaries orient the microtubules.
The long-range collective orientational ordering of the array 
then propagates this information over the entire surface.

I would propose an alternative variant of the boundary-conditions
explanation of orientation bias; my version depends on the 
azimuthal periodicity but not on the end caps.  
We know, from simulations or common sense, that before the system coarsens into 
a single ordered domain, the orientational correlations extend
much farther in the direction parallel to the microtubule orientation
than transverse to it.  
When that orientation happens to be azimuthal.
the correlated domain has its best chance of growing all the wall
around the cell and finding itself, thereby locking the orientation.
(This presumes the cell's circumference is smaller than its length.)
Finally, yet another possible explanation of the transverse orientation 
bias is that the microtubules independently sense the curvature direction
of the membrane (see Sec.~\ref{sec:kinked-attachment}).

The least clear aspect of array alignment is 
is why that orientation {\it deviates} from transverse 
with a definite handedness, on average, as observed experimentally.
Let's start with a coarse-grained description at the level of
a continuum theory or a mean-field theory.
Imagine the array's collective order parameter angle 
$\alpha(t)$ {\it precesses} at some average rate~\cite{LR-Landau},
   \beq
      d\alpha/dt = \omega_0.
   \label{eq:mt-precess}
   \eeq
In fact, video imaging showed local rotations of
array domains in elongating plant cells~\cite{chan-mt-video} 
and Wasteneys also speculated there could be a global
precession rate~\cite{wasteneys-ambrose}.
A dynamic rotation is also suggested by morphologies of some
plant cells that have multiple layers of cellulose fibers,
each rotated relative to the one underneath~\cite{preston82}.
%%%%%%%%%%%%%%%%%%%%%%%%%%%%%%%%%

Meanwhile, whatever its origin, the transverse orientation
bias will appear as a competing term added into \eqr{eq:mt-precess}.
Then the combination of the precession force 
and the transverse bias yields a final orientation
deviating in a particular sense from transverse, by
an angle $\alpha_0$ dependent on the relative
strength of the two factors~\cite{LR-Landau}.

\subsubsection{Realizing chiral bias in microtubule simulations?}

Let us now step down in level and consider explicitly the
stochastic dynamics of many microtubules that follow the above
listed event rules.  Chiral bias can be incorporated in the dynamics
via either kind of event that is an interaction:
\begin{itemize}
\item{\it Branching}:
the branching angle distribution, of a new microtubule
relative to the old one, is asymmetric between left
and right.
%%%%%%%%%%%%%%%%%%
\item{\it Collision}:
the outcome of an mt-mt collision depends on which 
side the growing mt is impinging from [Rule (iv)].
\end{itemize}
Since the constituents are chiral, it is
a trivial observation that no exact symmetry 
dictates that either process has an exact L/R symmetry.
Then a nonzero precession rate is {\it generically} expected;  
however, it might be imperceptibly small in practice, and
will be suppressed once orientational long range order 
extends all the way around the cell.

The only simulation that has produced a helical bias is
\cite{eren-mt-sim}; that was accomplished by combining
a chiral bias in the branching rates with an 
ad-hoc simulation condition, in which the rules were
changed partway through the run.  This change probably
realized the competition with a transverse orientation
bias as I suggested just above, with that bias effectively 
being provided by open boundary conditions representing the cell caps
in this simulation. Subsequently, Ref.~\cite{deinum-mt-sim} (their
Sec. 3.4) also simulated the effects of chiral branching bias 
under various conditions, and observed precession but did not 
quantify it; they concluded its effects are too weak to account
for the reorientations seen in Ref.~\cite{chan-mt-video}.
It remains to be checked whether a collision bias can 
have a stronger effect~\cite{shen-henley}.

\subsection{Molecular level: mechanisms for precession?}
\label{sec:mt-precession-molecular}

Let us step down again to the molecular level:
here we have stronger reasons to predict 
precession of the array based on either of the 
possible aligning mechanisms, as a handed bias 
of the branching is {\it observed}, and a bias of
the collisions can be argued theoretically.

\subsubsection{Branching mechanism}
\label{sec:mt-precession-branching}

A subtlety of the experimental data is that
although Ref.~\cite{chan-mt-data} (page 2300)
reported an equal likelihood of positive and negative branching angles
$\theta_b$, they also reported 
a statistically significantly difference
in the {\it distributiona} of angles.
Just as the effectiveness of branching for aligning the array
is roughly $\overline{\cos 2\theta_b}$ as derived from a
mean-field theory~\cite{hawkins-mt-math,deinum-mt-sim},
so the effectiveness for precession should be
$\overline{\sin 2\theta_b}$.  This quantity was estimated
based on the histograms in Ref.~\cite{chan-mt-data},
figures 2C and 3E, and indicates a significant bias~\cite{shen-henley}.
The sign is $\overline{\sin 2\theta_b}>0$, indicating a tendency
to precess CCW, that is $\omega_0 >0$ in \eqr{eq:mt-precess}.

It is expected that one of the ``membrane associated proteins'' 
(MAP) mediates the branching, by binding to the existing mt and nucleating
the new one.  
Since the mt consists of $\sim 13$ equivalent protofilaments,
the MAP could in principle bind at any azimuthal angle around the 
mt; presumably this orientation determines the direction in which 
the nucleated mt emerges.  The observed left/right differences in branching 
mean that the nearby membrane conditions the nucleated mt's direction;
presuming the latter is determined by the MAP orientation, it means that
is constrained by the membrane.  This is an example of the fundamental 
symmetry considerations of Sec.~\ref{sec:question-assumptions}; it may
be rather obvious, but it logically leads to an interesting biological 
prediction: there must exist some physical interaction between the MAP 
and the membrane -- an interaction which has not yet been identified
experimentally. By ``physical interaction'' I do not necessarily mean 
that the MAP binds to the membrane or associates with a membrane-bound 
protein: since microtubules in an array are at most 20 nm from the membrane,
steric hindrance -- the MAP bumping into the membrane in some orientations --
would suffice to constrain the angles of nucleated microtubules.
In any case, whatever the exact details, the branching mechanism is clearly of the 
``anchoring'' type, in the dichotomy of Sec.~\ref{sec:classify-screw}.

\subsection{Microtubule screw-ratchet mechanism}
\label{sec:mt-screw-ratchet}

Now I turn to collisions.  It is well known that filaments
 that are polymerizing in a non-equilibrium fashion can exert forces 
on (say) a perpendicular wall via the ``Brownian ratchet'' 
mechanism~\cite{brownian-ratchet}.
Repeatedly, the wall thermally fluctuates away from the tip,
this sterically allows another monomer to be added, which
like a strut prevents the wall from returning to its original
position.  Similar Brownian notions 
were used to estimate the probability for one growing
microtubule to cross over or under another one~\cite{allard-mechanical}.
These estimates did not take into account the closeness of the
membrane.

In fact, the microtubule adds tubulin units in a left-handed spiral fashion.
Therefore, by a brownian ratchet mechanism, it will exert forces
much like a left-handed screw  being screwed into a solid.
If it approaches another mt from the right, that screw path first encounters
the blocking mt as the screw is turning downwards, and hence tends to
lift  the incoming mt upwards -- towards the membrane, where it tends
to get sterically blocked and start depolymerizing.  On the other hand,
if the new mt comes in from the left, the screw path of polymerization
will push it downwards where there is more room, and the crossover
probability should be larger.
Evidently, in the dichotomy of Sec.~\ref{sec:classify-screw},
this collision mechanism is of the ``screw'' type.

The microtubules spiral counter-clockwise (looking in the direction
of growth), so it is the microtubules incident from the {\it left}
that tend to get pushed downwards and cross over successfully.
It follows that this mechanism predicts a sign of the precession 
$\omega<0$, i.e. clockwise.  In the experiment, 2/3 of the 15 cells 
shown in Fig.~1 of Ref.~\cite{chan-mt-video} showed a clockwise 
tendency, but that imbalance is not statistically significant,
and was not commented by the authors.

\subsubsection{Estimate of transverse ratchet force}

I will make a quantitative estimate whether the ratchet force is strong
enough to displace a microtubule far enough so as to cross over another one.
It is estimated that, for physiological concentrations of tubulin,
the available Brownian ratchet force is $\sim 30$pN; however, a lower
limit is implied by buckling.  The separation $\ell$ between points where the mt 
is anchored to the membrane is $\ell\sim 3\mu$m ~\cite{allard-mechanical}. 
Assuming a typical bending modulus $B \approx 20$ pN $\mu$m$^2$ for the microtubule, 
the buckling force is $B/D^2\sim 2 pN$. 
%%%%%%%%%%%%%%%%%%%
%%%%%%%%%%%%%%%%%%%
The unit added to a protofilament in each step is 8 nm long, the
mt diameter is $2 R=35$ nm, and there are 13 protofilaments.  
Also, each turn around adds 1.5 units in length.
From this
it can be figured out that the tangent vector along the
helix of successive addition points has longitudinal component
$\sim 1.5(8{\rm nm})/(2 \pi R) \approx 0.12$.  It follows that the transverse
component of the available ratchet force is $(0.12)(2 {\rm pN})\sim 0.2$pN.

What are we asking this force to do?  In the time that the growing microtubule 
would grow the radius of the barrier microtubule, it has to be displaced
sideways by its own radius so as to cross over.  A typical growth 
speed is $v_g \approx 3.5 \mu$m/min, so this time is 
$t \approx (R/v_g) \approx 0.3$s, and the displacement
needed is $\Delta z \approx R$.

The response of the microtubule is determined by a combination of
viscous drag and the microtubule bending modulus: a tranverse force 
$\Ftip$ applied on the tip as a sudden step causes an initial displacement at the
tip which spreads backwards.  The drag on a rod of length $L$ moving 
at velocity  $v$ in a fluid of viscosity $\eta$,
analogous to Stokes's formula for a sphere, is given by
$\Fdrag/L = \zeta_\perp$  where
where 
   \beq
       \zeta_\perp \equiv \frac{4\pi \eta }{-0.9 + \ln (L/R)}.
   \eeq
Taking $L\to\ell$ and assuming $\eta=10^{-2}$Pa s (ten times the
viscosity of water), we get $\lambda \approx 0.06 pN s/\mu{\rm m}^2$.
A proper treatment gives a differential equation for the displacement
$z(x,t)$ along the microtubule:
   \beq
        \frac{\partial z(x,t)}{\partial t} = 
         - \frac{B}{\zeta_\perp} \frac{\partial^4 z(x,t)}{\partial x^4},
   \label{z-t-PDE}
  \eeq
as is discussed in Ref.~\cite{wiggins-viscoelastica}.
For our purposes, however, dimensional analysis 
will suffice.
We expect the tip displacement to grow with time $t$ as
   \beq
        \Delta z(t) \sim \Ftip \Big(\frac {t^3}{B \lambda^3}\Big)^{1/4}.
   \label{z-t-dimensional}
   \eeq
Setting $\Delta z=R=17.5$nm and $t=0.3$s as argued above, we finally
get $\Ftip=0.01$pN as the minimum force needed,
modulo the unknown coefficient of order unity in \eqr{z-t-dimensional}.
This is just 1/20 of the estimated transverse ratchet force, suggesting 
that force is strong enouigh to induce a significant bias in the
crossover probability, depending which side the growing tube advances from.
Specifically, 
as noted in Section \ref{sec:mt-precession-branching}, 
since the barrier microtubule is separated from the membrane,
by $< 20$ nm $< 2R$, the crossover chance should be
seriously reduced for a displacement in this direction.
%%% $\lesssim$ requires {amsmath} package

\subsection{Kinked-attachment mechanism}
\label{sec:kinked-attachment}

As already mentioned in Sec.~\ref{sec:plant-cell}
the greatest unsettled question about the plant microtubule
array, from a physicist's perspective, is how does it (tend to)
align almost transverse to the cell's long axis, but slightly
helical?  Sec.~\ref{sec:plant-cell} suggested mechanisms whereby
this depended on the orientational ordering of the microtubule array.
Here I want to explore the alternative possibility that the
orientation bias depends on {\it local} properties of the
membrane, specifically its curvature.

\subsubsection{Molecular level argument}
\label{sec:kinked-attachment-molecular}

Consider the chiral properties of gram-negative 
bacteria like E. Coli: they are believed to be determined
by an actin homologue called MreB, which tends to wind around 
the cell membrane in a spiral, and seems to template the
bacterial cell wall, made of peptidoglycans, somewhat as a 
plant cell's microtubules template its cellulose cell wall.
But the MreB consists of a single bundle, not an array, so
its orientation cannot be imputed to a collective effect.
The membrane-binding type mechanism I suggest here could apply 
either to MreB or to plant microtubules.

\begin{figure}
\includegraphics[width=0.87\linewidth]{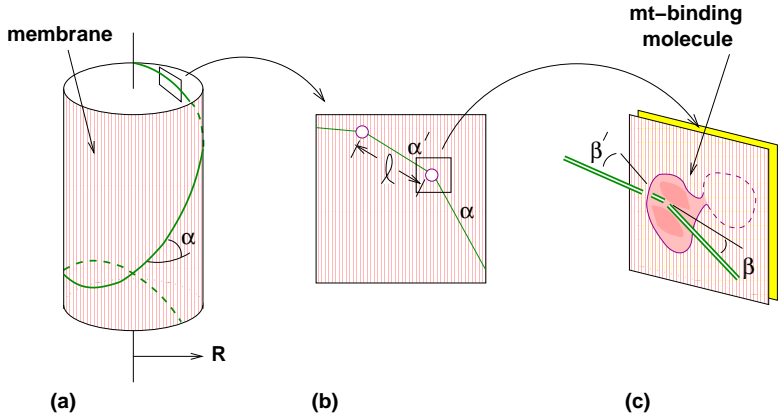}
\caption{Kinked-attachment mechanism for chirality
of a single filament, e.g. a microtubule,
(a) a cylindrical cell with the filament
(heavy line) wrapped around it in a helix parametrized
by a pitch angle $\alpha$;  (b) zoom to a portion
of the membrane:  the path actually
consists of straight segments between membrane-bound
molecules (circles) that clasp the filament, separated
by a  typical spacing $\ell$; the incoming angle $\alpha$ need not match 
the outgoing angle $\alpha'$ (defined with respect to the
filament's polarization) (c) on the molecular scale,
the filament is clasped in a membrane-bound molecule,
such that the filament is also bent in the direction
normal to the membrane; here $\beta$ and $\beta'$ are
the incoming and outgoing angle between filament and
membrane.}
\label{fig:kinky-bound}
\end{figure}

The filament's persistence length exceeds the cell diameter; 
to make it wrap around the cell [see Fig.~\ref{fig:kinky-bound}(a)],
bending forces must be exerted on it by (I suggest) the same molecules
that bind the filament. 
Thus we get binding points with a typical spacing $\ell$ (modifiable
under the cell's control),
with a {\it kink} at each binding point
and straight segments in between them.
We can characterize
the spatial relation of the filament to the binding protein by two angles
($\beta,\alpha)$ for the incoming line, and $(\beta',\alpha')$ for the
outgoing line [see Fig.~\ref{fig:kinky-bound}(b,c)], where $\beta$ or $\beta'$
is the angle to the membrane plane, and $\alpha$ or $\alpha'$ is
the direction within the membrane plane.  If the binding points are
equally spaced then $\beta=\beta'$ is constrained by the cell being 
a cylinder; for simplicity we assume $\beta=\beta'$ (and small) henceforth.
A nonzero $\beta$ corresponds to filament curvature $2\beta/\ell$,
which in turn is $2\cos^2\alpha/R$, where $R$ is the cell's radius.
Thus
   \begin{equation}
    \beta(\alpha) = \frac{\ell \cos^2 \alpha}{2R}.
   \label{eq:alpha-beta}
   \end{equation}
%%% for Fig.~\ref{fig:kinky-bound}
A certain set of angles $\{ \alpha, \beta, \alpha', \beta'\}$ minimizes
the binding molecule's (free) energy; any deviation from these angles has
an elastic cost $U$ which is quadratic in deviations.   Locally the physics
is symmetric around the membrane normal direction, so the in-layer angles
enter only through the difference $\Delta\alpha\equiv \alpha'-\alpha$,
but there is no reason to expect symmetry around 
$\Delta\alpha=0$. Thus the form is
   \begin{equation}
     U= \half K_{11} (\Delta \alpha)^2 + K_{12} \Delta\alpha (\beta-\beta_0)
     + \half K_{22} (\beta-\beta_1)^2.
   \label{eq:U-mt-binding}
   \end{equation}
Say we are given the incoming angle $\alpha$ [which by \eqr{eq:alpha-beta}
constrains $\beta$]: the optimum value  of $\Delta \alpha$, corresponding to
mechanical equilibrium, is
$\Delta alpha = -(K_{12}/K_{11}) (\beta-\beta_0)$.
Rewriting in differential form and substituting from 
\eqr{eq:alpha-beta}, the end result is
    \beq
        \frac{d\alpha}{ds} = - \frac{K_{12}}{K_{11}} \frac{\ell}{4R} 
                            \Big(\cos 2\alpha-\cos 2 \alpha_0\Big),
    \label{eq:alpha-Dalpha}
    \eeq
where $\alpha_0$ is defined by $\beta(\alpha_0)\equiv \beta_0$.
Eq.~\eqr{eq:alpha-Dalpha}
is the desired formula for expressing the selection of a helix with
a particular angle $\alpha_0$.
The basic idea is that, due to the generic cross term in 
Eq.~\eqr{eq:U-mt-binding}, the sense of the filament's in-plane
bending is controlled by whether its out-of-plane curvature
is greater or less than a threshold; 
that out-of-plane curvature in turn is controlled by the pitch 
angle $\alpha$.  As a result, 
a filament at angle $\alpha_0$ turns neither to the right nor to
the left. Filaments at other angles bend within the plane of the
membrane, until they are at angle $\alpha_0$.
All this depends on having $K_{12}>0$ in 
\eqr{eq:U-mt-binding}; the other sign implies a positive feedback
on $\alpha-\alpha_0$, in which case the favored angle would be either
$0^\circ$ or $90^\circ$.

%%%%%%%%%%%%%%%%%%%%%%%%%%%%%%%%%%%%%%%%%%%%%%%%%%%%
\section{Brain handedness, chiral crawling and actin pitch-change}
\label{sec:actin-motility}

Up to now, all our examples of chirality in animals
were in the context of development.  
I now turn to some less-developed or recent emerging
stories that involve
cell growth and/or motility and/or adhesion.

Perhaps the most studied asymmetry is that of 
the human brain, and perhaps the mechanism of 
its implementation is the least understood.
Nerve cells migrate far in the developing brain.
After stopping they extend long ``neurites''
(meaning axons or dendrites) which will form 
synapses to interconnect with other nerve cells. 
This growth is not so different from locomotion: 
either phenomenon involves chemotaxis and cytoskeletal 
elements pushing out the membrane.

A difficulty is that, if brain symmetry is broken
at a late stage when the tissue is three-dimensional,
what specifies locally the directions ``$\xaxis$'' and ``$\zaxis$'' 
which enter into a right-hand rule?
In most of the other examples of L/R phenomena surveyed in
this paper, the cells form a two dimensional layer, 
with the upper and lower sides inequivalent,
so that we get one axis for free.
Is it possible that brain asymmetry instead arises as
a very early mechanism while the embryo is roughly
two dimensional? Could this be related to the 
L/R mechanism which Levin~\cite{levin-birthdefects}
claims exists in vertebrates as an alternative to
the cilia-driven flow?
%%% left-right anomalies seen in twins

\subsection{Chirality in nerve cell growth}
\label{sec:neurites}

One possible microscopic basis for brain handedness is 
the noticeably chiral growth of the neurites.
Originally this was observed as a tendency of
neurites grown on a flat substrate to 
curve rightwards with a radius
of curvature $\sim 1$ mm \cite{agranoff-neurites,tamada-neurites} 
or much less ~\cite{romijn-neurites,levitan-neurites}.
Recently, in a three-dimensional gel,
it was shown the neurite tip's motion could be idealized 
as a right-handed helix of radius $\sim 2\mu$m and pitch $\sim 20\mu$m 
\cite{tamada-neurites}.
%%% $50 \mu$m in \cite{romijn-neurites},
%%% $30\mu$m in \cite{levitan-neurites}.
This was shown to depend on the processive form of
Myosin V, which steps in a 
spiraling fashion~\cite{ali-helical}.

All in all, this top-level story seems very reminiscent of 
plant roots, (see Section~\ref{sec:macro-root}, below). 
The turning on a flat substrate being analogous to plant roots 
that do the same, or that exhibit root skewing (see 
Section~\ref{sec:macro-root}); the chemotaxis of the
neurite toward its target is analogous to the gravitropism
(downward bias) of the plant roots.  Further understanding
would benefit from mechanical studies of the neurite elasticity
and forces exerted during growth, perhaps presenting artificial
mechanical barriers as the three-dimensional root experiments of
Ref.~\cite{jesse-roots} or the microfabricated channels in the
cell swarming experiments (Sec.~\ref{sec:mammalian-channels},
and \ref{sec:bacteria-flagellated}, below).

Detailed studies of axon growth over time showed that the 
tip behaves reminiscent of the famous chemotaxis of the
bacterium {\it E. Coli}. It follows a fixed bearing, modulo 
inherent angular fluctuations, for a long time;  then with,
a time constant $\sim 4$ hours, it executes a turn to a new
bearing -- which is chirally biased to be $\sim 30^\circ$ 
to the left of the old one (Ref.~\cite{sanjana-thesis}, figures 4C,
5B, and 6B).

In the case of {\it E. Coli}, the turn is executed by tumbling,
so one wonders  if the neurite's turn might be analogously 
accompanied by some sharp looping twist, which could be 
revealed experimentally by detailed microscopy at the points of turning.
Whereas {\it E. Coli} tumbles many times so as to randomize
its new direction, the speculated looping of the neurite would
have to be less than a full turn so as to keep its correlation
with the preceding direction. As in all the other examples,
since an up/down asymmetry is a prerequisite to distinguishing
left and right within the planem, an experimental suggestion
is to vary that asymmetry, perhaps by growing neurites confined
on both sides.  A related question is to compare conditions
leading to different macroscopic radius of curvature: is this
simply a change in the frequency of reorientation events, 
or a change in the degree of bias in each event?

On the molecular level, the next experimental step would be to 
address the necessity of microtubules, actin, and myosin for the
bias in the turning.  It would also be of interest to survey
mutations known to affect brain laterality, identify any that
relate to cytoskeletal genes, and then see whether the chirality
of neurite growth is disrupted in the mutants.

\subsection {Actin pitch-change mechanism}
\label{sec:actin-pitch}

One way to generate torsion from actin filaments is to take advantage 
of the inherent elasticity of a chiral filament, which by symmetry
should include a bilinear cross-coupling between the longitudinal
strain and the torsion.
That term has been measured in DNA~\cite{ENS-twist-stretch,bustamonte-twist-stretch}.
but not yet, it seems, in actin (where it may be less important:
see \cite{kamien-twist-stretch}, footnote 3).

Consider a polymerizing pseudopod, made of an actin network that
pushes the cell membrane as it polymerizes.  Presumably, each new
actin filament is under a small tension or compression. 
But as the actin network exerts a forwards force 
on the membrane, it must develop internal longitudinal stresses,
which due to the cross-coupling induce a torque of the same sense
in each filament.  Since these filaments are cross-linked they 
cannot unwind freely, and the network has a macroscopic torsion.
Inside an approximately planar cell crawling on a surface, this would
tend to twist the pseudopod out of that plane.  Granting that the
``upper'' (free) and ``lower'' (surface-bound) sides of the cell are
inequivalent, the twist converts this to a left-right inequivalence.
If the actin network experiences a different chemical environment as 
it matures, so as to change the preferred pitch, we get the same outcome.
One would thus predict that eukaryotic cells, in chemotaxis,
would walk systematically offset from the gradient direction
of the attractant chemical.

No significant offset in the angle of chemotaxis has never been reported
~\cite{arrieu-compass}.  Recently, however, a chiral offset
was found in animal cells when a spatially 
{\it uniform} concentration of attractant is increased.
Such a step in time is known to induce a spontaneous polarity, 
as defined by the nucleus-centrosome axis, and it turns out 
that the new axis is mainly $\sim 90^\circ$ to the left
of the old one~\cite{left-polarizing}.  
However, this behavior was shown to depend 
on {\it microtubules}, which was not found to date in any
of the other animal-cell stories.  It would obviously make
sense to check experimentally for the involvement of myosin 
and/or of actin polymerization. Since different cell types
display opposite handedness in channels 
(Sec.~\ref{sec:mammal-channels-micro}, below), 
one would also be curious whether they also display
opposite rotations of the polarization axis.

\subsection{Collective chiral crawling}
\label{sec:mammalian-channels}

Very recently,
massive chirality was found in mature, motile mammalian
cells.  Cells were packed into micropatterned planar
channels, and the layers near the walls were found to
move with opposite senses on either side, like automobile
traffic on a road; whether the cells ``drive'' on the
right or the left depends on the cell type, and even
on whether they are cancerous~\cite{Wan-Vun-channels}.
This only occurs at high packing densities of the cells,
so it must be a collective effect of their interactions,
presumably related to cell-cell adhesion.  

\subsubsection{Collective level: chiral medium?}

Strikingly, the cells showed a biased motion in annuli as well
as in straight channels, but not in a space shaped like an 
open disk.  If the bias were attributed to a direct interaction
of cells sensing the wall~\cite{Wan-Vun-channels}, 
flow along the edge would be seen 
even in the dilute limit of random-walking single cells, and it 
would not depend on the opposite boundary; that is what swarming
bacteria do (Sec.~\ref{sec:bacteria-flagellated}.  Instead,
I conjecture that the mammalian cells are actually reacting 
to each other, developing a chiral stress.   
In the middle of the swarm, this stress is uniform so its
effects cancel and nothing is seen; but adjacent to walls, 
it is unbalanced and drives the flow.

Thus the collective story seems fully analogous to that of
the actomyosin layer in Sec.~\ref{sec:actomyosin-screw}.
In place of filaments, the active units here are cells,
but the essential fact is the same: nearby units have
an interaction which tends to orbit them in a particular
sense.  I would speculate that the coarse-grained flow 
of a cell population is described by the 
same continuum theory presented in 
Sec.~\ref{sec:actomyosin-continuum}.

One likely experimental test of this, 
as in Sec.~\ref{sec:actomyosin-continuum},
would be to analyze video images by cross-correlating the 
density fluctuation gradients with velocity fluctuations.  
The  images show several ``convection rolls'' spaced around the
annulus, so the first check would be to see if the cores of these 
rolls have a different cell density.
This system offers additional opportunities to control
density gradients, in the form of the wall constraints
imposed by the fabricated  channel geometry.  
If the continuum theory correctly predicts the absence of rotation
in an open disk, it should also predict the differences between
a variety of other shapes of the accessible region.

As noted above, I am suggesting that the cells
are {\it not} recognizing the walls, but only each other;
the walls act as purely mechanical boundary conditions.
That could be tested experimentally by modifying the wall surfaces:
the prediction is that treating them with signals
would have less effect than changing their coefficient of friction.

\subsubsection{Cell level story: inter-cellular forces 
and actin polymerization?}
\label{sec:mammal-channels-micro}

What future experiments might uncover the rules
governing the motions of individual cells and interacting pairs?
One obvious fact, following the dictum in Sec.~\ref{sec:x-z-cross},
is that the cells must know the difference between up and down
in order to tell the difference between left and right.  
Presumably gravity is a minor effect, so if the cells were
equally in contact with the ``floor'' and ``ceiling'' of
their channels, there should be no right/left bias.  
Actually, a cancellation in that case follows from basic
symmetry so it is predicted for any mechanism; 
but manipulations of the up/down differences could reveal 
which kind of difference is salient for the chiral flow, 
which would in turn be a
clue for the mechanism at the next lower level.

The nicest way to unravel the cell interactions would be
to use a mixture of two cell types.~\footnote{
I am grateful to Victor Luria for this suggestion.}
One would independently vary (i) how they recognize 
each other (ii) whether they are right or left biased cell types.
By examining the dynamics of individually imaged cells,
one can distinguish between various candidate interactions
which would all give the same collective behavior in a
uniform system.

At the molecular level, experiment
already showed that the mechanism 
does not require microtubules, but does
depend on actin polymerization,
as it is suppressed by the polymerization inhibitors 
latrunculin and cytochalasin~\cite{Wan-Vun-channels}.
Furthermore it is found {\it not} to depend on myosin 
(Ref.~\cite{Wan-Vun-channels}, Fig.~S5).
Hence the mechanism must be different from the actomyosin
story of Sec.~\ref{sec:actomyosin}; instead, it presumably 
involves the basic cell motility, which certainly depends 
on actin polymerization.
Thus, one possibility is to impute the mechanism to something
like the actin pitch-change mechanism of Sec.~\ref{sec:actin-pitch}.

Since I argued the handedness in this system lies in cell-cell interactions,
that might suggest testing for the involvement of cadherins, 
involved with cell-cell adhesions.  
Cadherins were actually implicated
in a mechanism for {\it Drosophila} (but with immotile cells in that
case:  see Sec.~\ref{sec:drosophila}).

\subsection{Chirality in flies}
\label{sec:drosophila}

In the highly studied development of the fruit fly {\it Drosophila},
a small (non-functional?) chirality is manifested as 
looping of their guts and 
genitals~\cite{speder-drosophila-orig,hozumi-drosophila-orig,speder-noselli-review0}.
This appeared at a late, many-cell stage and 
was shown to depend on the motor protein 
``unconventional'' myosin type I D, as well as myosin I C;
Mutants in which either I D is nonfunctional, or else I C is overexpressed,
show a reversed handedness; thus it appears the respective
myosins are antagonists.
I include this topic because a new experiment~\cite{drosophila-tissue} 
found clues which may be sufficient information for us to start 
(speculatively) modeling that system.

%% In this experiment the myosin I D is spatially associated with the 
%% adherens junction, i.e. with how cells stick to each other.

\subsubsection{Top level story: tissue shear}

\begin{figure}
\includegraphics[width=0.87\linewidth]{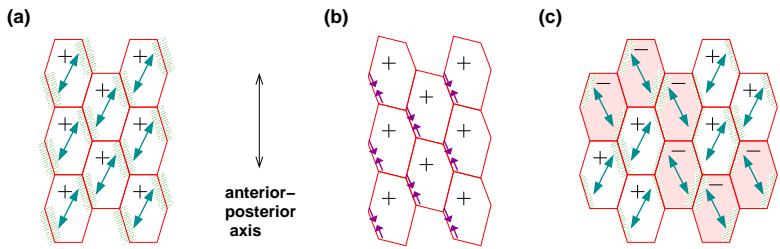}
\caption{Left-right asymmetry in hindgut epithelial 
tissue of {\it Drosophila}.
 (inspired by Ref.~\cite{drosophila-tissue}.
In this cartoon, emphasizing the asymmetries, 
cells are represented as elongated hexagons,
omitting the irregularities of the actual tissue 
(shown in Ref.~\cite{drosophila-tissue}, Figures 2 and 3).
(a) In the wild-type (cells with $+$) cadherin ($D$E-Cad, shown 
by hatching) is
observed to be localized on the walls tilted leftwards from 
the anterior-posterior (AP) axis; this is attributed to
a hypothetical myosin-based transport of vesicles 
along the arrowed directions.  (b) In the final tissue,
the orientation of walls has a systematic bias (i.e. the
tissue is sheared); this is attributed to contraction
caused by the cadherin (opposing arrows).
(c) In a tissue mixing wild-type ($+$) cells with mutant 
($-$, shaded) cells, the cadherin density along each cell wall
is the sum of the densities attributable to the respective
adjoining cells.}
\label{fig:canted-brickwork}
\end{figure}

Namely, it found that the epithelial (tissue layer) cells
lining the gut exhibit a form of chirality that I have not
mentioned up to now.
%%%%%%%%%%%%%%%
%%%%%%%%%%%%%%%
The packing of the cells has a preferred axis (anterior to posterior)
and then this tissue gets sheared, such that cell walls transverse to that 
axis are biased to tilt at a nonzero angle, as idealized in 
figure~\ref{fig:canted-brickwork}.
This macroscopic shear breaks right-left and up-down mirror symmetries,
but does not in itself break inversion symmetry, and thus does not
distinguish left and right directions.
If, though, there were a prexisting anterior (head-wards) polarization,
such a shear would generate a rightwards polarization.
%%% 

Ref.~\cite{drosophila-tissue} 
also found that cadherins  -- membrane proteins that mediate
cell-cell adhesion and modify the stress tensor -- preferentially
localize on cell boundaries with (two-dimensional) normal vectors
oriented in the first and third quadrants, roughly the (1,1)
vector in the plane, as represented in 
Fig.~\ref{fig:canted-brickwork}(a).
In the inversion mutant
this preference is reversed; in a mosaic tissue where both 
types of cells are mixed randomly, each category of cell boundary 
has a cadherin preference which is the mean of that for
the cell types on either side.   As recognized in Ref.~\cite{drosophila-tissue},
the extra cadherins on boundaries normal to the (1,1) direction drive a
cortical tension and contraction normal to (1,1), as shown in
Fig.~\ref{fig:canted-brickwork}(b).
When  -- as in the hindgut --  the tissue
is rolled into a tube, this shear drives a spiral twist of the tube, as
verified in Ref.~\cite{drosophila-tissue} by a toy simulation.
Thus, the top level of the L/R mechanism -- represented by continuum
elasticity --  is qualitatively understood.

Furthermore, when the investigators engineered a mixture of the 
wild-type $(+)$ and the reversal mutant $(-)$ cells, they showed 
that cadherin was distributed on the $(++)$ type walls with the 
same bias as in wild type, but the 
$(- -)$ walls were biased the opposite way, 
and the $(+-)$ walls were not significantly biased,
as shown in Fig.~\ref{fig:canted-brickwork}(c).

\subsubsection{Cell-level mechanism: biased transport}
\label{sec:drosophila-cell}

At the molecular level,
myosins I C and I D are known to transport vesicles 
(small membrane bags) containing proteins -- perhaps the cadherins
-- along actin fibers.  Thus we need a mechanism for biased transport. 
A possibly useful fact is that the mutants with inverted handedness do 
{\it not} have a reversal in the physical bias of the transport, 
but rather in the sign of the signal which gets transported
~\cite{drosophila-tissue}.  Ref.~\cite{hozumi-drosophila-orig} already
recognized that either the actin filaments are aligned exactly 
along the anterior/posterior axis but the myosin transport is
biased, or else the fibers themselves are turned. I will
describe slightly more explicit candidate mechanisms
corresponding to each of thes possibilities.

The first candidate mechanism is ``asymmetric transport'' 
which I worked out in Sec. 3 of Ref.~\cite{LR-Landau} 
for purely pedagogical purposes,  whereby 
a filament array oriented bidirectionally along the 
A/P axis {\it without}  L/R bias, does transport vesicles
with a diagonal bias.  The first condition for this
is easy enough:  a helical bias in the
myosin I D step, similar to the myosin II that is better
known and discussed in Sec.~\ref{sec:actomyosin-screw}, above.
%%%%%%%%%%%%%%%%%%%%
%%%%%%%%%%%%%%%%%%%%
The second condition is that the fibers are flush against
a membrane which sterically contrains the tendency of the
vesicles to spiral around the actin as they are pulled
by myosin I D.  For this mechanism to give a non-cancelling
result, the A/P polarized array of actin filaments must sit on 
one large surface of the cell (say the upper one) but not the other.

As an alternative candidate mechanism, one could posit a 
large scale flow of a cortical actomyosin layer as in 
Sec.~\ref{sec:actomyosin}, which will tend to run counter-clockwise
next to the cell boundaries, as we look down on the upper cell faces.
If the transporting actin array has its endpoints embedded 
in that cortical layer, the array's orientation 
tends to precess with that layer, just
as I posited the microtubule array does in plant cells
(Sec.~\ref{sec:plant-cell}).  Just as in that story, 
if this precession is combined with a competing pull towards
ia particular axis (here the A/P axis), the result will be
a static orientation with a mean tilt.  But as with the 
asymmetric-transport mechanism, the key question is:
why either the top or the bottom face of the cell dominates
over the other one?  I would suggest that further experiments
focus on the up/down difference.

Based on the above speculations, it seems fairly plausible that 
the mechanism in {\it Drosophila} -- whatever it is -- belongs
to the ``linear response'' category of Sec.~\ref{sec:SSB},
and not to the more common ``symmetry breaking'' category.

%%%%%%%%%%%%%%%%%%%%%%%%
%%%%%%%%%%%%%%%%%%%%%%%%

%%%%%%%%%%%%%%%%%%%%%%%%%%%%%%%%%%%%%%%%%%%%%
\section{Bacteria}
\label{sec:bacteria}

Bacteria exhibit a number of handed phenomena that are naturally
handled under the same framework as multicellular organisms.  
Of course, there is not (necessarily) an inter-cellular level here, 
but even in single cells one needs the same multiple levels of 
description (see below).  An important motivation for me to include
this (very incomplete) sketch about bacteria is that 
they are the subject of the best-developed existing theory that
applies to left/right phenomena
the sort of multiscale approach that I am espousing in this paper.
Such treatments are usually set up at the level of mechanics 
or fluid dynamics.

\subsection{Flagellated bacteria}
\label{sec:bacteria-flagellated}

It is already well-known that in flagellated bacteria, such as 
{\it E. Coli}, there is a hydrodynamic interaction of the
flagella with nearby walls that drives a biased turning of
the bacterium's path around the axis of the wall normal~\cite{swim-in-circles}.
It was found that {\it E. Coli} in microchannels swim on the right
as {\it isolated} cells,~\cite{Ecoli-right-side}
thus indicating a relatively trivial non-cooperative 
reason for the observed chiral swarming motions~\cite{Bacteria-SWARM},
as opposed to the collective origin I conjectured for the chiral
swarming of mammalian cells (Sec.~\ref{sec:mammalian-channels}).

Flagellated bacteria are also long known to form chiral colony
patterns on agar gels~\cite{ben-jacob} and this has been understood
theoretically in terms of their single-cell dynamics 
on a surface~\cite{ben-jacob-theory}. 

Later, a continuum theory~\cite{herbie-levine} 
was written down aiming to model this growth, 
using a precessing orientation field similar 
in meaning to my $\theta(\rr)$ from the plant 
microtubule arrays (Sec.~\ref{sec:mt-plants}).
Simulations of these equations produce 
feather-like chiral patterns similar to the 
bacterial colonies; tip-splitting instabilities
are central to this pattern formation.

\subsection{Non-flagellated bacteria and their cytoskeleton}
\label{sec:bacteria-MreB}

I now turn to bacilli (rod shaped bacteria) without 
heavy cell walls such as {\it B. subtilis} or
{\it Caulobacter}.
Independent of whether they have flagella,
many species adopt naturally handed spiral shapes;
mutants do in the cases of 
{\it B. Subtilis}~\cite{mendelson-1976}
or {\it E. Coli}~\cite{varma-young}
The external shape presumably depends on the organization
of the layer of peptidoglycans that form a cell wall
outside the membrane.  It is believed that the actin homolog MreB, 
which forms a single helical bundle just inside 
the membrane, right-handed in {\it B. subtilis}~\cite{MreB-right-handed}
or left-handed in {\it E. Coli}~\cite{MreB-wall-helical},
directs the wall synthesis~\cite{MreB-wall-helical,figge-MreB-wall}.
Thus the story is reminiscent of plant cells,
in which the cellulose synthesis (forming the cell wall)
is directed by the microtubules, which organize helically
inside the membrane, as I discussed at length in Sec.~\ref{sec:mt-plants}.

One reason to bring up this system is that it demonstrates
that it not necessary to form a collective state in order to
bias the sense of the helix, which suggested to me the likelihood
of something like the ``kinked-attachment'' mechanism of 
Sec.~\ref{sec:kinked-attachment-molecular}.
Also, we see it is not essential to have inter-cellular interactions
in order to encounter multiple levels of description:
filament polymerization dynamics at the bottom, perhaps 
a statistical description of the cell wall and its continuum theory 
as an intermediate level, and mechanics determining external shape 
at the top level.

Existing theory is often based on the notion of twist-to-writhe
conversion: if you twist a rod, differential geometry shows 
the strain is partially reduced when the centerline becomes helical, 
and this may be the mechanically stable state.
Ref.~\cite{wolgemuth} addresses how an existing spiral in the
cylindrical cell can, by elongating and pressing against the
end caps, drive the external shape to a spiral.
Ref.~\cite{MreB-wall-helical} contains a different theory
based, more plausibility, on the helical  correlations built
into the peptidoglycan wall,  because its new strands are
nucleated from the MreB filament.

Another macroscopic -- or intercellular -- level handedness is 
the long-known supercoiling of the strands of cells formed
when mutant {\it B. Subtilis} cells divide without 
separating~\cite{mendelson-supercoil}. 
This hints at a more microscopic origin similar to
the division-twist posited in snail embryose
(Sec.~\ref{sec:mollusc-division-twist}). 
There is an analogy to the micro level in which
each linked cell in a strand corresponds to a monomer 
in a cytoskeletal filament.

Turning to the molecular level, but still in the spirit of the 
twist-to-writhe conversion, Ref.~\cite{andrews-arkin-MreB} 
has explained why the MreB filament forms a helix within the cell.
They treat the MreB as a ribbon that, like DNA, has an inherent twist, 
and that attaches to the membrane with its flat side parallel.
Notice that -- unlike most other mechanisms mentioned at this level --
this is a pure mechanical equilibrium theory of passive
elements, that does not depend on polymerization or any other
driven behavior.
It has been shown that MreB forms a two-strand helix 
%%%%%%%%%%%%%%%%%%%%%%
\OMIT{These strands are related by 2-fold rotation, whereas in DNA 
the strands are related by inversion. (Wingreen, personal communication.)}
%%%%%%%%%%%%%%%%%%%%%%
-- so it really is a ribbon -- furthermore MreB binds directly to
the membrane without need for a linker~\cite{MreB-membrane-lowe}
Hence, for the case of MreB Ref.~\cite{andrews-arkin-MreB}
appear to have the correct molecular picture
as opposed to the ``kinked-attachment'' one in 
Sec.~\ref{sec:kinked-attachment}. 
%%%%%%%%%%%%%%%%%%%%%%%%%%%%%%%%%%%%%%
\OMIT{Is the differential geometry -- how the ribbon responds to 
variations in the cylinder diameter, or to a boundary condition 
-- that constrains its end to a different angle from $\alpha$
-- different from that one?}

\section{Discussion and conclusions}

In this speculative review, I have asserted it is
helpful to approach left/right asymmetries
in disparate biological systems in parallel,
with a common framework, rather than in a
compartmentalized fashion.  To illustrate this,
I discussed, or at least mentioned, the following examples:
\begin{itemize}
 \item
node cilia in vertebrates (Sec.~\ref{sec:cilia-flow});
\item 
actomyosin, in snail embryos and also in {\it C. elegans}
(Sec.~\ref{sec:actomyosin})
\item
the microtubule layer in plant cells (Sec.~\ref{sec:mt-plants});
\item
neurites, perhaps ultimately the human brain
(Sec.~\ref{sec:neurites});
\item
collective swarming by motile mammalian cells
(Sec.~\ref{sec:mammalian-channels})
\item
chirality of cell packing in {\it drosophila} 
(Sec.~\ref{sec:drosophila}).
\item
bacteria with the actin homolog MreB (Sec.~\ref{sec:bacteria});
\end{itemize}
(Sec.~\ref{sec:macro-root} also very briefly mentioned the spiral
growth of the fungus {\it Phycomeces}; this would be very
attractive to develop as another example.)

I have talked about general principles, and presented several
examples, of how chiral information gets from the molecular
scale to the macro scale, so that (almost) all individuals 
break left/right symmetry in the same ways.
Each story naturally unfolds on multiple length scales:
there is always an intracellular
story and an organism level story, and sometimes more than two levels. 
So far, all the mechanisms imagined, and all
those observed, involve the cytoskeleton (microtubules or actin filaments)
at the molecular level.

Just how can the molecular chirality affect larger
scale dynamics?
In Sec.~\ref{sec:classify-screw}, I identified 
two general ways.  The first is a ``screw'' mechanism, 
meaning motion {\it along} a long helical fiber
gets converted into rotation {\it around} the fiber,
due e.g. to a motor's walking, the collisions of a helically polymerizing
microtubule, or to the fiber's twist/extension cross-coupling.
(This feels attractive to physicists since it uses the translational 
invariance of the long fiber and its inherent chirality.)

The second mechanism is the ``membrane-anchoring'' mechanism: some
membrane-bound molecule also specifically binds the fibers in a 
particular orientation.  (This scenario will feel more familiar
to biologists.) The membrane-anchoring mechanism furnishes a hint
for genetic or protein-expression studies as to the existence, or the
importance, of such a protein.

@1
Symmetry principles (Sec.~\ref{sec:question-assumptions})
called for,  in any L/R mechanism and at any level, 
not only a chiral element but also two axes.
But in the majority of examples given above, where we have a glimmer
of understanding, we find a screw mechanism giving rise to a
it only produced a handed rotating tendency in the dynamics of this array:
a ``relative'' chirality, in the categories of Sec.~\ref{sec:classify-screw}:
This was the case for the shearing actomyosin layer,
(Sec.~\ref{sec:actomyosin-continuum}), or the 
orientationally precessing microtubule arrays.
(Sec.~\ref{sec:plant-cell}).
In these cases, the conjectured cell-level mechanism 
is easier to implement since it requires only {\it one} pre-existing axis 
-- the membrane normal or cell layer normal, depending on what level we are at.
To produce a macroscopic handedness, this rotating dynamics 
must occur in  the context of some larger-scale nonuniformity --
the furrow between dividing animal cells, or the curvature 
directions and end caps of a plant cell --

It is also striking that the top level story
usually involves mechanics.  This may be ascribed to how
L/R asymmetry is represented: we ultimately need 
``positional information'', meaning each cell has access
to a signal which tells which side of the L/R line it is on,
but the lower level mechanisms typically produce local 
{\it polarization} of the cells.
Mechanics is a convenient way to convert that to 
a macroscopic distortion, an operation which is 
a bit like inverting the derivative in calculus
(as I commented in Sec.~1.3.3 of Ref.~\cite{LR-Landau}).
The other way to convert orientational information to
positional information is by using polarized, driven 
transport to build up a gradient of some signal.
The only instances of that are transport by the
surrounding fluid, at the top level of the 
nodal-flow story of vertebrates (Sec.~\ref{sec:cilia-flow}) 
or in the likely vesicle transport, in the cell-level
story of flies (Sec.~\ref{sec:drosophila-cell}).

I have given, where I could, a few experimental preductions,
or I pointed out experiments which would clarify a mechanism.
The recurring theme in these suggestions, in all the different
examples,  correlates with the above observation that 
chirality develops in layers.  Thus, we need experiments
to unravel what aspect of the up/down asymmetry of such layers
controls the left/right asymmetry within them: any such 
discoveries will be clues for the left/right mechanism.

It is left for future research to turn all these ideas into
quantitative estimates, a necessary condition before any
physics may be considered as completed.

%%%%%%%%%%%%%%%%%%%%%%%%%%%%%%%%%
%%%%%%% END STUFF   %%%%%%%%%%%%%%%
%%% \begin{acknowledgments}
{\sl Acknowledgments}
\LATER{(JSP template fails?)}
I thank 
S. Grill, 
T. Hashimoto, 
F. J\"ulicher, 
M. Levin,
V. Luria,
N. Sanjana,
S. Seung,
E. D. Siggia, 
G. O. Wasteneys,
N. Wingreen,
R. Chachra, 
and J. X. Shen, 
for discussions  and communications
(and the last two for collaborations).
This work was supported by the U.S. Dept. of
Energy, grant DE-FG-ER45405.
%%% \end{acknowledgments}

%%%%%%%%%%%%%%%%%%%%%%%%%%%%%%%

\end{document}